\documentclass[preprint,journal]{vgtc_arxiv}                

\ifpdf
  \pdfoutput=1\relax                   
  \usepackage{graphicx}                
  \DeclareGraphicsExtensions{.pdf,.png,.jpg,.jpeg} 
\else
  \usepackage{graphicx}                
  \DeclareGraphicsExtensions{.eps}     
\fi%

\graphicspath{{figures/}{pictures/}{images/}{./}} 

\usepackage{microtype}                 
\PassOptionsToPackage{warn}{textcomp}  
\usepackage{textcomp}                  
\usepackage{mathptmx}                  
\usepackage{times}                     
\usepackage{amssymb}
\usepackage{cite}                      
\usepackage{tabu}                      
\usepackage{booktabs}                  
\usepackage{algorithmic}
\usepackage{algorithm}
\usepackage{pbox}
\usepackage{calrsfs}
\usepackage{amsmath}
 \usepackage{enumitem}

\DeclareMathOperator*{\argmin}{arg\,min}
\DeclareMathAlphabet{\pazocal}{OMS}{zplm}{m}{n}

\onlineid{1010}

\vgtccategory{Research}
\vgtcpapertype{Technique \& Algorithm}
\ieeedoi{10.1109/TVCG.2019.2934256}
\title{Progressive Wasserstein Barycenters of Persistence 
Diagrams\vspace{-.75ex}}

\author{Jules Vidal, Joseph Budin, and Julien Tierny}
\authorfooter{
\item
J. Vidal, J. Budin, J. Tierny are with Sorbonne Universit\'e and CNRS 
(LIP6). 
E-mail: \{jules.vidal, joseph.budin, julien.tierny\}@sorbonne-universite.fr.
}

\shortauthortitle{Biv \MakeLowercase{\textit{et al.}}: This is the title}

\abstract{
This paper presents an efficient algorithm for the progressive approximation of
Wasserstein barycenters of persistence diagrams, with applications to the
visual analysis of ensemble data. Given a set of scalar fields, our approach
enables the computation of a persistence diagram which is representative of the
set, and which visually conveys the number, data ranges and saliences of the
main features of interest found in the set. Such representative diagrams are
obtained by computing explicitly the discrete Wasserstein barycenter of the set
of persistence diagrams, a notoriously computationally intensive task.
\julienRevision{\EDIT{,
  for which we introduce}{. To do so, we build upon existing work on the
  computation of Fr\'echet means and on fast approximations of Wasserstein
distances for diagrams to design}}
{In particular, we revisit efficient algorithms for Wasserstein 
distance approximation \cite{Bertsekas81, Kerber2016} to 
extend previous work on barycenter estimation \cite{Turner2014}. We 
present}
a new fast algorithm, which progressively
approximates the barycenter by iteratively increasing the computation accuracy
as well as the number of persistent features in the output diagram. Such a
progressivity drastically improves convergence in practice and allows to design
an interruptible algorithm, capable of respecting computation time constraints.
This enables the approximation of Wasserstein barycenters within interactive
times. We present an application to ensemble clustering where we 
revisit the
$k$-means algorithm to exploit our barycenters and compute, within execution
time constraints, meaningful clusters of ensemble data along with their
barycenter diagram. Extensive experiments on synthetic and real-life data sets
report that our algorithm converges to barycenters that are qualitatively
meaningful with regard to the applications, and quantitatively comparable to
previous techniques, while offering an order of magnitude speedup when run
until convergence (without time constraint). Our algorithm can be trivially
parallelized to provide
additional speedups in 
practice
on standard workstations.
We provide a lightweight C++ implementation of our approach that can 
be used to reproduce our results.
\vspace{-.75ex}
}

\CCScatlist{ 
 \CCScat{K.6.1}{Management of Computing and Information Systems}%
{Project and People Management}{Life Cycle};
 \CCScat{K.7.m}{The Computing Profession}{Miscellaneous}{Ethics}
}

\newcommand{\domain}{\mathcal{M}}
\newcommand{\range}{\mathbb{R}}
\newcommand{\criticalIndex}{\mathcal{I}}
\newcommand{\persistence}{\mathcal{P}}
\newcommand{\persistenceDiagram}[1]{\mathcal{D}(#1)}
\newcommand{\projection}{\Delta}
\newcommand{\pointMetric}[1]{d_{#1}}
\newcommand{\liftedMetric}[1]{\widehat{\pointMetric{#1}}}
\newcommand{\wasserstein}[1]{W_{#1}}
\newcommand{\diagramSpace}{\mathbb{D}}
\newcommand{\isovalue}{w}
\newcommand{\sublevelset}[1]{#1^{-1}_{-\infty}}
\newcommand{\balancedPersistenceDiagram}{\mathcal{D}'}
\newcommand{\benefit}{\beta}
\newcommand{\price}{p}
\newcommand{\bidValue}{v}
\newcommand{\priceDiff}{\delta}
\newcommand{\overallCost}[1]{\widehat{\wasserstein{#1}}}
\newcommand{\auctionTolerance}{\gamma}
\newcommand{\diagramSet}{\mathcal{F}}
\newcommand{\diagramBarycenter}{\mathcal{D}^*}
\newcommand{\diagram}{\mathcal{D}}

\newcommand{\needsVerification}[1]{\textcolor{red}{#1}}
\renewcommand{\needsVerification}[1]{\textcolor{black}{#1}}
\newcommand{\persistenceThreshold}{\rho}
\newcommand{\threadNumber}{n_t}
\newcommand{\timeConstraint}{t_{max}}
\newcommand{\criticalPointMixing}{\lambda}
\newcommand{\geometricLifting}{\alpha}
\newcommand{\persistenceScaling}{\tau}
\newcommand{\removable}[1]{\textcolor{purple}{#1}}
\renewcommand{\removable}[1]{}
\newcommand{\mixRef}{\cite{Turner2014}+\hspace{-.05ex}\cite{Kerber2016}}

\newcommand{\EDIT}[2]{\textcolor{black}{#2}}
\newcommand{\julienRevision}[2]{\textcolor{black}{#2}}

\newcommand{\todo}[1]{\textcolor{red}{TODO : \emph{#1}}}
\renewcommand{\todo}[1]{}

\newcommand{\imageCaption}[1]{
\vspace{-3ex}
\caption{#1}
\vspace{-1ex}
}

\newcommand{\finalVersionSpace}[1]{\vspace{#1}}

\teaser{
  \vspace{.75ex}
  \centering
  \includegraphics[width=\linewidth]{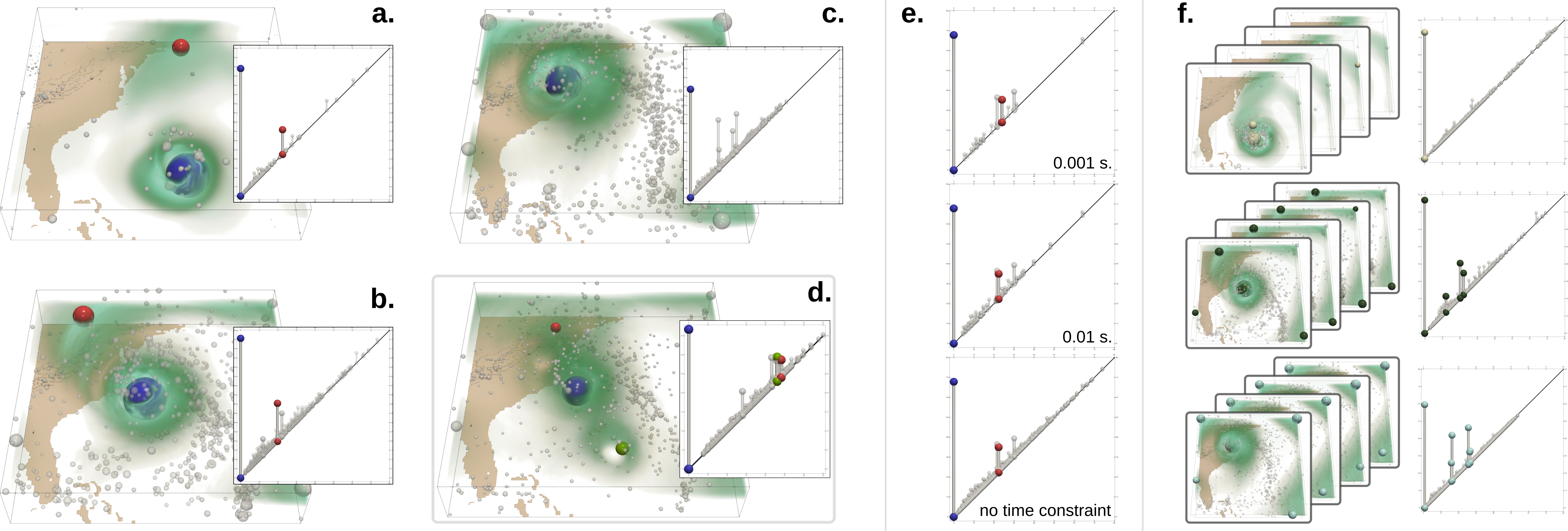}
  \imageCaption{
The  Persistence
diagrams of \EDIT{3}{three} members (a-c) of the Isabel ensemble (wind velocity) 
concisely and visually encode the number, data range and salience of the 
features of interest found in the data
(eyewall and region of high speed wind, 
blue and red in 
(a)).
In these diagrams, features with a \emph{persistence} smaller than 10\% 
of the function range or 
\julienRevision{due to}{on}
the boundary are shown in transparent white. 
The pointwise mean for these \EDIT{3}{three} members (d)
exhibits \EDIT{3}{three} salient
interior features (due to 
distinct eyewall locations,
blue, green and red), 
although the diagrams of the input members only report \EDIT{2}{two} salient 
interior features at most, located at drastically different data ranges (the 
red feature is further down the diagonal in (a) and (b)). The Wasserstein 
barycenter of these \EDIT{3}{three} diagrams (e) 
provides a more representative view of the features found in this ensemble, as 
it reports a feature number, range and salience that better matches the input 
diagrams (a-c). Our work introduces a progressive approximation algorithm for 
such 
barycenters, with fast practical convergence. Our framework supports computation 
time constraints (e) which enables the approximation of Wasserstein 
barycenters within interactive times. We present an application to the 
clustering of ensemble members based on their persistence diagrams ((f), 
lifting: \needsVerification{$\geometricLifting = 0.2$}), which 
enables the visual exploration of the main trends of features of interest found 
in the ensemble. 
\vspace{1ex}
}
	\label{fig:teaser}
}

\vgtcinsertpkg

\usepackage{xcolor}
\begin{document}



\renewcommand{\sectionautorefname}{Sec.}
\renewcommand{\subsectionautorefname}{Sec.}
\renewcommand{\figureautorefname}{Fig.}
\renewcommand{\equationautorefname}{Eq.}
\renewcommand{\tableautorefname}{Tab.}
\newcommand{\algorithmautorefname}{Alg.}

\firstsection{Introduction}

\maketitle

In many fields of science and engineering,
measurements and simulations are 
core tools for the understanding of complex physical systems.
However, given the 
geometrical complexity of the resulting data,
interactive exploration and analysis can be challenging
for users. This motivates the design of expressive data abstractions, capable 
of concisely capturing the main features of interest in the data, and of 
visually conveying that information to the user, quickly and effectively.

\begin{figure}
  \centering
  \includegraphics[width=.95\columnwidth]{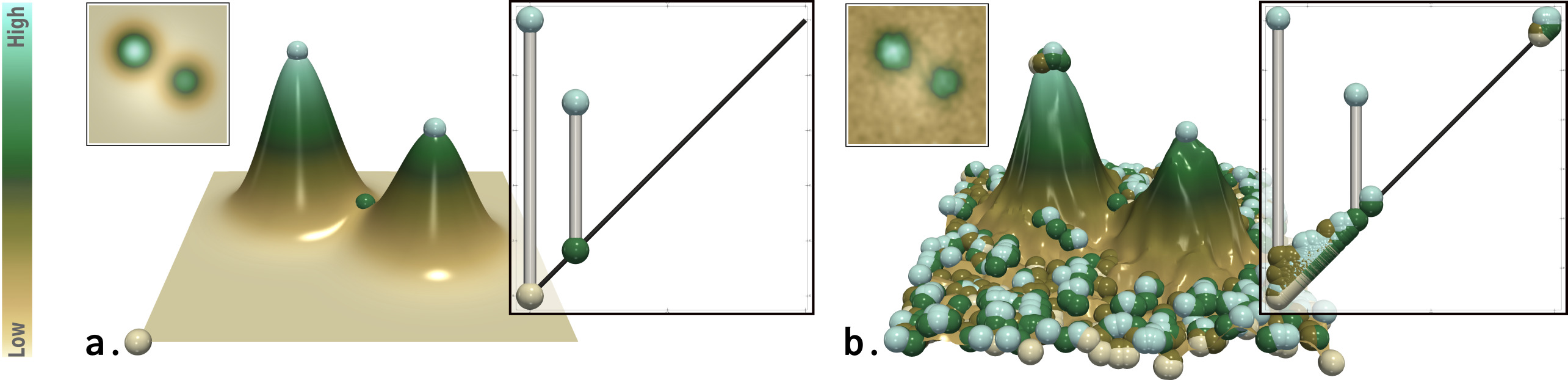}
  \vspace{1ex}
  \imageCaption{Critical points (spheres, light brown: minima, light green: 
maxima, 
other: saddles) and persistence diagrams of a clean (a) and noisy (b) 2D scalar
  field. From left to right : 2D data, 3D terrain visualization, persistence 
diagram.
  In both cases, the two main hills are clearly represented by 
  salient persistence pairs in the diagrams.
  In the noisy diagram (b), small pairs near the diagonal correspond
  to noisy features in the data.}
  \label{fig_persistenceDiagram}
\end{figure}

In that regard, Topological Data Analysis (TDA) \cite{edelsbrunner09} 
has demonstrated  over the last two decades its utility to support interactive 
visualization tasks \cite{heine16}. In the applications, it robustly and 
efficiently captures in a generic way the features of interest in scalar data. 
Examples of successful applications include 
turbulent combustion \cite{laney_vis06, bremer_tvcg11, gyulassy_ev14}, material 
sciences \cite{gyulassy_vis07, gyulassy_vis15, favelier16, roy19}, 
nuclear energy \cite{beiNuclear16},
fluid dynamics \cite{kasten_tvcg11}, 
medical imaging \cite{beiBrain18},
chemistry \cite{chemistry_vis14, harshChemistry} or 
astrophysics \cite{sousbie11, shivashankar2016felix} to name a few. 
An appealing aspect of TDA is the ease it offers for the translation of
domain-specific descriptions of features of interest into topological terms. 
Moreover, to distinguish noise from features, 
concepts from Persistent Homology 
\cite{edelsbrunner02, 
edelsbrunner09} provide importance measures, which are both theoretically well 
established and meaningful in the applications.
Among the existing abstractions, such as contour trees \cite{carr00}, 
Reeb graphs \cite{pascucci07, biasotti08, tierny_vis09}, or Morse-Smale 
complexes \cite{gyulassy_vis08, gyulassy_vis14, Defl15,gyulassy19}, the persistence 
diagram \cite{edelsbrunner02} has been extensively studied. In 
particular, its conciseness, stability \cite{cohen-steiner05} and expressiveness 
make it an appealing candidate for data summarization tasks. For instance, its 
applicability as a concise data descriptor has been well studied in machine 
learning \cite{ReininghausHBK15, Carriere2017, rieck_topoInVis17}.
In visualization, it provides visual hints about the number, data range
and salience of the features  of interest 
(\autoref{fig_persistenceDiagram}), which helps users 
visually 
apprehend the complexity of their data. 

In practice, modern numerical simulations are subject to a variety of input 
parameters, related to the initial conditions of the system under study, as 
well as the configuration of its environment. 
Given recent advances in hardware
computational power,
engineers and scientists can now densily sample the space of these input 
parameters, 
in order to better quantify the sensitivity of the system.
For scalar variables, this means that the data which is considered 
for visualization and analysis is no longer a single field, but a 
collection, called \emph{``ensemble''}, of scalar fields representing the same 
phenomenon, under distinct input conditions and parameters. In this context, 
extracting the global trends in terms of features of interest in the ensemble 
is a major challenge.

\begin{figure}
  \centering
  \includegraphics[width=.95\columnwidth]{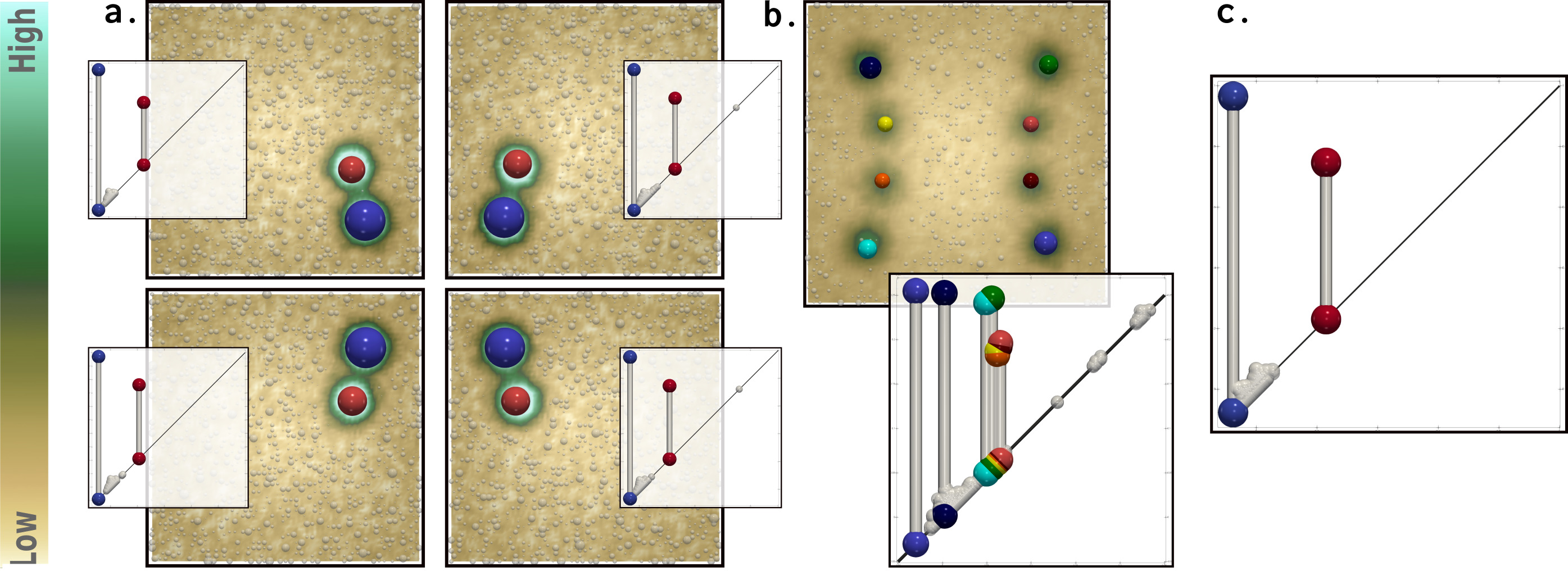}
  \vspace{1ex}
  \imageCaption{Synthetic ensemble of a pattern with 2 gaussians and additive 
noise 
(a). The persistence diagram of the pointwise mean (b) contains 8 highly 
persistent features although each of the input ensemble members contain only 2 
features. The Wasserstein barycenter  (c) provides a 
diagram which is representative of the set, with a feature number, range 
and salience which better describes the input ensemble (2 large features).
\vspace{-3ex}
}
\label{fig_motivation}
\end{figure}

Although it is possible to compute a 
persistence diagram
for each member of an 
ensemble, 
in particular in-situ  \cite{insitu, AyachitBGOMFM15}, 
this process 
only shifts the problem from 
the analysis of an ensemble of scalar fields to an ensemble of 
persistence diagrams.
Then, given such an ensemble of diagrams,
the question 
of estimating 
a diagram
which is \emph{representative} of the set naturally 
arises, as such a representative diagram could visually convey to the users 
the global trends in the ensemble in terms of features of interest.
For this, naive strategies could be considered, such as estimating the 
persistence diagram of the  mean of the ensemble of scalar fields. However, 
given the additive nature of the pointwise 
mean, this yields a persistence diagram with an incorrect number of 
features (\autoref{fig_motivation}), which is thus not representative of any of 
the diagrams of the input 
scalar fields. To address this issue, a promising alternative consists in 
considering the 
\emph{barycenter} of a set of diagrams, given a distance metric between them, 
such as the so-called \emph{Wasserstein} metric \cite{edelsbrunner09}, hence 
the term 
\emph{Wasserstein barycenter}. 
For this, an algorithm has been proposed 
 by Turner et al. \cite{Turner2014}. However, it is based on an 
iterative procedure, 
for which each iteration relies itself on a demanding optimization problem 
(optimal 
assignment in a weighted bipartite graph \cite{Munkres1957}), which makes it 
impractical for real-life datasets. 

This paper addresses this problem by introducing a fast 
algorithm for the approximation of discrete Wasserstein barycenters of 
ensembles of 
persistence diagrams. 
We designed our approach by revisiting the 
core routines involved in this optimization problem with appropriate 
heuristics, motivated by practical observations.
A unique aspect of our approach is its progressive 
nature: the computation accuracy and the number of features in the 
output diagram are progressively increased along the optimization.
This specificity has two main practical advantages. First, 
this progressivity drastically accelerates convergence in practice. Second, it 
enables to formulate an \emph{interruptible} algorithm, capable of producing 
meaningful barycenters while respecting computation time constraints. This 
latter advance is particularly useful, both in an interactive exploration 
setting (where users need rapid feedback) and in an in-situ context (where 
computation ressources need to be carefully allocated).
Extensive experiments on synthetic and real-life data report that our algorithm 
converges to barycenters that are qualitatively meaningful 
for the 
applications, while offering 
an order of magnitude speedup over the fastest combinations of existing 
techniques. Our 
algorithm is 
\EDIT{trivially}{easily} parallelizable, which provides 
additional speedups on
commodity workstations.
We illustrate the utility of our approach 
by introducing
a clustering algorithm adapted from $k$-\emph{means} which 
exploits 
our progressive Wasserstein barycenters. This application enables a meaningful 
clustering 
of the members based on their features of interest, 
within computation time constraints, and provides informative 
summarizations of the global trends of features found in the 
ensemble.
%

\subsection{Related work} 
\label{sec_relatedWork}
The literature related to our work can be classified into three main 
categories, reviewed in the following:
\emph{(i)} uncertainty visualization, \emph{(ii)} ensemble visualization, and 
\emph{(iii)} persistence diagram processing.


\noindent
\textbf{Uncertainty visualization:} The analysis and visualization of 
uncertainty  in data is a notoriously challenging problem in the visualization 
community \cite{uncertainty_survey1,unertainty_survey2, uncertainty1, 
uncertainty2, uncertainty3,
uncertainty4}.
In this context, the data variability is explicitly modeled by an estimator of 
the probability density function (PDF) of a pointwise random variable.
Several  representations
have been proposed to visualize the related data uncertainty,
either
focusing on the entropy of the random variables 
\cite{uncertainty_entropy}, on their 
correlation \cite{uncertainty_correlation}, or gradient
variation \cite{uncertainty_gradient}. When geometrical constructions are
extracted from uncertain data, their positional uncertainty has to be assessed.
For instance, several approaches have been
presented for level sets, under various interpolation schemes and  PDF models
    \cite{uncertainty_isosurface1,uncertainty_isosurface2,
    uncertainty_isosurface3, uncertainty_isosurface4, uncertainty_isocontour1,
uncertainty_nonparam,uncertainty_isocontour2, uncertainty_interp,athwale19}. 
Other approaches addressed critical point positional uncertainty, either for 
Gaussian \cite{liebmann1,otto1,otto2,petz} or uniform distributions 
\cite{gunther,bhatia,szymczak}. 
In general, visualization methods for uncertain data are specifically designed 
for a given distribution model of the pointwise random variables (Gaussian, 
uniform, etc.). This challenges their usage with ensemble data, where PDF 
estimated from empirical observations can follow an arbitrary, unknown model. 
Moreover, most of the above techniques do not consider multi-modal PDF models, 
which is a necessity when several distinct trends occur in the ensemble.

%
%


\noindent
\textbf{Ensemble visualization:}
Another category of approaches has been studied to specifically visualize 
the main trends in ensemble data. In this context, the data 
variability is directly encoded by a series of global empirical observations 
(\emph{i.e.} the members of the ensemble). Existing visualization techniques 
typically 
construct geometrical objects, such as level sets or streamlines, for each 
member of the ensemble. Then, given this ensemble of geometrical objects, 
the question of estimating an object which is \emph{representative} of the 
ensemble naturally arises.
For this, several methods have been proposed,
such as spaghetti plots \cite{spaghettiPlot} in the case of level-set
variability in weather data ensemble \cite{Potter2009,Sanyal2010}, or box-plots
for the variability of contours \cite{whitaker2013} and curves in general
\cite{Mirzargar2014}. For the specific purpose of trend variability analysis,
Hummel et al. \cite{Hummel2013} developed a Lagrangian framework for
classification in flow ensembles. Related to our work, clustering
techniques have been used to analyze the main trends in ensembles of
streamlines\cite{Ferstl2016} and isocontours \cite{Ferstl2016b}. However,
only few techniques have focused on applying this strategy to topological 
objects.
Overlap-based heuristics have been investigated to estimate a representative 
contour tree from an ensemble \cite{Wu2013ACT,Kraus2010VisualizationOU}.
Favelier et al. \cite{favelier2018} introduced an approach to analyze
critical point variability in ensembles. It relies
on the spectral clustering of the ensemble members according to their
\textit{persistence map}, a scalar field defined on the input geometry which 
characterizes
the
spatial layout of persistent critical points. However, the clustering stage of 
this approach takes as an input a distance matrix between the persistence 
maps of all the members of the ensemble, which requires to mantain them all in 
memory, which is not conceivable for large-scale ensembles counting a high 
number of members. Moreover, the clustering itself is performed on the spectral 
embedding of the persistence maps, where distances 
are loose approximations of the
intrinsic metric between these objects.
In contrast, the clustering application described in this paper directly 
operates on the persistence diagrams of the ensemble members, whose memory 
footprint is orders of magnitude smaller than the actual ensemble data. This 
makes our approach more practical, in particular in the perspective of an 
in-situ computation \cite{insitu} of the persistence diagrams. 
Also, it 
is built on top of 
our 
progressive algorithm for Wasserstein barycenters,
which focuses on the Wasserstein metric between diagrams.
Optionally, our work can also
integrate the spatial layout of critical points by considering a 
geometrically lifted version of this metric \cite{soler2018}.
\noindent
\textbf{Persistence diagram processing:}
To define the barycenter of a set of persistence diagrams, a metric \emph{(i)} 
first needs 
to be introduced to measure distances between them. Then, barycenters 
\emph{(ii)} can be 
formally defined as minimizers of the sum of distances to the set of diagrams.
%

\vspace{-.5ex}
\emph{(i)} The estimation of distances 
between topological abstractions has been a long-studied problem, in particular 
for similarity estimation tasks. 
Several heuristical  approaches have been documented for the fast estimation of 
structural similarity 
\cite{hilaga:sig:2001,thomas14,SaikiaSW14_branch_decomposition_comparison,ThomasN13}.
More formal approaches have studied various metrics 
between topological abstractions, such as Reeb graphs \cite{bauer14,BauerMW15} 
or
merge trees \cite{BeketayevYMWH14}.
For persistence diagrams, the \emph{Bottleneck} \cite{cohen-steiner05} and 
\emph{Wasserstein} distances \cite{monge81, Kantorovich, edelsbrunner09}, 
have been
widely studied, for instance 
in 
machine learning \cite{cuturi2013} and adapted for
kernel based methods \cite{ReininghausHBK15, Carriere2017, 
rieck_topoInVis17}.
%
%
%
%
%
%
%
The numerical computation of the 
Wasserstein distance between two persistence diagrams requires to solve an 
optimal assignment problem between them. The typical methods 
for this
are the exact Munkres algorithm
\cite{Munkres1957}, or an auction-based approach \cite{Bertsekas81} that
provides an approximate result with improved time performance. 
Kerber et al.
\cite{Kerber2016} specialized the auction algorithm to the case of
persistence diagrams and showed how to
significantly improve performances by leveraging adequate data structures for 
proximity queries.
Soler et al. \cite{soler2018} introduced a fast extension of 
the Munkres approach, also taking advantage of the structure of 
persistence diagrams, in order to solve
the assignment problem in an exact and efficient way, using a reduced, sparse 
and unbalanced
cost matrix. 

\vspace{-.5ex}
\emph{(ii)} Recent advances in optimal
transport \cite{cuturi2013}
enabled the practical
resolution 
of transportation problems between continuous quantities \cite{cuturi2014, 
solomon2015}. 
These
methods have been successfully applied to persistence diagrams
\cite{lacombe2018}.
However, this application requires to represent the input diagrams as 
heat maps of fixed resolution \cite{Adams2015}.
Such a rasterization can be interpreted as a pre-normalization of the 
diagrams, which can 
be problematic for applications where the considered diagrams have different 
persistence scales,
as typically found with time-varying phenomena 
for instance. 
%
Moreover, this approach does not explicitly produce a persistence diagram as 
an output, but a heat map of the population of persistence pairs in the 
barycenter. This challenges its usage 
for 
visualization applications, as the features of interest in the 
barycenter 
cannot be directly inferred from the barycenter heat map. In contrast, our 
approach produces explicitly a persistence diagram as an output, from which
the geometry of the features (their number, data ranges and salience)
can be directly visually inspected. Moreover, such an explicit representation 
also enables the efficient geometrical lifting of the Wasserstein metric 
\cite{soler2018}, which is relevant for scientific visualization applications 
but which would require to regularly sample a five dimensional space with 
heat map based approaches.
%
Turner et al. \cite{Turner2014} introduced an algorithm for the computation of 
a Fr\'echet mean of a set of persistence diagrams with regard to the 
Wasserstein metric. 
%
This approach provides explicit barycenters, which makes it appealing 
for the applications. However, its very high computational cost makes it 
impractical for real-life data sets. In particular, it is based on an iterative 
procedure, 
for which each 
iteration relies itself on $N$ optimal assignment problems \cite{Munkres1957} 
between persistence diagrams (for the Wasserstein distances), 
where $N$ is the number of members in the ensemble.
A naive approach to address this computational bottleneck would be to combine 
this method with the efficient algorithms for 
Wasserstein 
distances mentioned previously \cite{Kerber2016, soler2018}. However, as 
shown in \autoref{sec_results},
such \EDIT{a naive}{an} approach is still 
computationally expensive and it can require up to hours of computation
on 
certain data sets. 
In contrast, our algorithm converges in multi-threaded mode 
in a couple of minutes at most, and its progressive nature
additionally allows for its interruption within interactive times, while still 
providing qualitatively meaningful results.


\subsection{Contributions} This paper makes the following new contributions:
\vspace{-1.5ex}
\begin{enumerate}[leftmargin=1em]
\item{\emph{A progressive algorithm for Wasserstein
barycenters of persistence diagrams:}
\julienRevision{\EDIT{We present }{We extend existing work on Fréchet means of 
persistence
  diagrams \cite{Turner2014} and revisit the fast approximation of Wasserstein
distances \cite{Bertsekas81,Kerber2016} in order to provide }}
{We revisit efficient algorithms for Wasserstein 
distance approximation \cite{Bertsekas81, Kerber2016} 
in order 
to 
extend previous work on barycenter estimation \cite{Turner2014}. In 
particular, we introduce}
a new 
\julienRevision{algorithm}{approach}
based on a progressive
  approximation 
strategy, which iteratively refines both computation accuracy and output 
details. 
The persistence pairs of the input diagrams are 
progressively considered in decreasing order of persistence. 
This focuses the computation towards the most salient features of the 
ensemble,
while considering noisy persistent pairs last.
The returned barycenters are explicit and provide insightful visual hints 
about the features present in the ensemble.
\julienRevision{We show that our}{Our}
progressive strategy drastically accelerates 
convergence in practice, resulting in an order of magnitude speedup 
over
the fastest combinations of existing techniques. The algorithm is trivially 
parallelizable, 
which provides additional speedups 
in practice 
on standard workstations.
\julienRevision{\EDIT{}{Additionally, the progressivity allows to broaden the 
above algorithm to 
support computation time constraints, and produce barycenters accounting
for the main features of the data within interactive times.}}
{We present an \emph{interruptible} extension of our algorithm to support
computation time constraints. This enables to produce barycenters
accounting for 
the main features of the data within interactive times.}}

\vspace{-1.5ex}
\item{\emph{An interruptible algorithm for the clustering of persistence 
diagrams:}
We extend the above methods to revisit the
$k$-\emph{means} algorithm and 
introduce 
an interruptible 
clustering of persistence diagrams, which  is used 
for the visual analysis of the
global feature trends in ensembles.}
\vspace{-3.5ex}
\item{\emph{Implementation:} We
provide  a  lightweight  
C++ implementation of our algorithms
that can be used for reproduction purposes.}
\end{enumerate}

\section{Preliminaries} 
This section presents the theoretical background of our approach. It contains 
definitions adapted from the Topology ToolKit~\cite{ttk17}. It 
also provides, for self completeness, concise descriptions of the key 
algorithms 
\cite{Turner2014, Kerber2016} that our work extends. 
We refer the reader to Edelsbrunner and Harer \cite{edelsbrunner09} for an
introduction to  computational topology.

\begin{figure*}
\includegraphics[width=\textwidth]{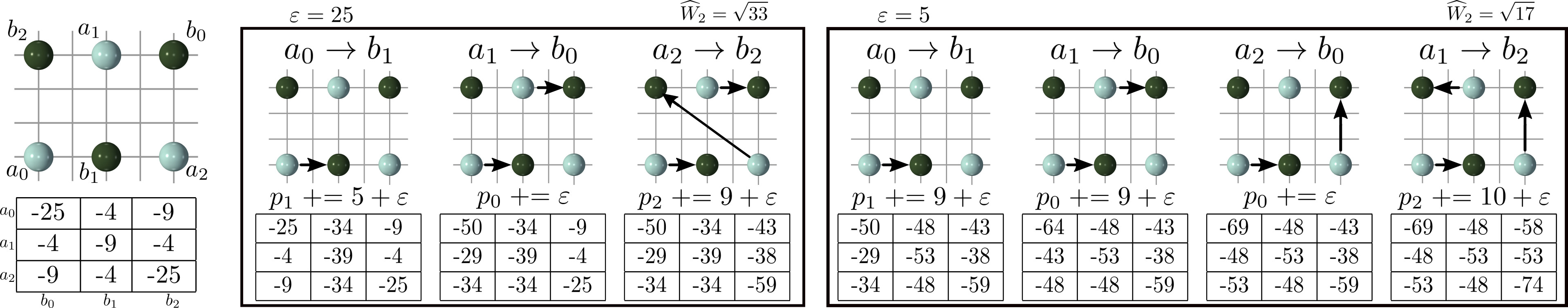}
\imageCaption{Illustration of the \emph{Auction} algorithm \cite{Bertsekas81} on 
a 2D 
 example for 
the optimal assignment of 2 point sets (light and dark green).
Boxes and columns represent \emph{auction rounds} and 
\emph{auction iterations} respectively.
Each matrix reports \needsVerification{the 
value $\bidValue_{a \rightarrow b}$ 
currently estimated by the bidder $a$ for the purchase of 
the object $b$}. 
Assignments are shown with black arrows. After the first 
round (left box), a sub-optimal assignment is achieved, which becomes optimum 
at the second round (right box). Note that bidders can steal
objects from each other within one auction round (iteration $3$, second 
round).}
 \label{fig_aucionExample}
\end{figure*}
\subsection{Persistence diagrams}

The input data
is an ensemble of $N$ piecewise linear (PL) scalar fields $f : \domain 
\rightarrow \range$ defined on  a PL $d$-manifold $\domain$, with
$d\leq3$ in our applications. 
We note $\sublevelset{f}(\isovalue)=\{p \in \domain~|
~f(p) < \isovalue\}$ the \textit{sub-level set} of $f$, namely the
pre-image of $(-\infty, \isovalue)$ by $f$. When continuously increasing 
$\isovalue$, the topology of 
$\sublevelset{f}(\isovalue)$ 
can only 
change at specific locations, called the \emph{critical points} of $f$. 
In practice, $f$ is enforced to be injective on the vertices of $\domain$ 
\cite{edelsbrunner90}, which are in finite number. This guarantees
that the critical points of $f$ are isolated and also in finite number. 
Moreover, they are also enforced to be non-degenerate, which can be easily  
achieved with local re-meshing \cite{edelsbrunner03}.
Critical points are classified according to their \textit{index} 
$\criticalIndex$ :
0 for minima, 1 for 1-saddles, $d-1$ for $(d-1)$-saddles, and $d$ for
 maxima.
%
Each topological feature of $\sublevelset{f}(\isovalue)$ 
(i.e. connected 
component, independent cycle, void)
can be 
associated with a unique pair of critical points $(c, c')$, corresponding to 
 its 
\emph{birth} and \emph{death}.
Specifically, the Elder rule 
\cite{edelsbrunner09} states that critical points can be arranged according to 
this observation in a set of pairs, such that each critical point appears in 
only one pair $(c, c')$ such that $f(c) < f(c')$ and $\criticalIndex{c} = 
\criticalIndex{c'} - 1$. Intuitively, this rule implies that if two topological 
features of  $\sublevelset{f}(\isovalue)$ (\emph{e.g.} two connected 
components) meet at a critical point $c'$, the \emph{youngest} feature 
(i.e. created last) \emph{dies}, favoring the \emph{oldest} one (i.e. created 
first). Critical point pairs can be visually represented by the 
\emph{persistence diagram}, noted $\persistenceDiagram{f}$, which embeds each 
pair to a single point in the 2D plane at coordinates $\big(f(c), f(c')\big)$, 
which respectively correspond to the birth and death of the associated 
topological feature. The \emph{persistence} of a pair, noted $\persistence(c, 
c')$, is then given by its height $f(c') - f(c)$.
It describes the lifetime in the range of the 
corresponding topological feature.
The construction of the persistence diagram is often restricted to a specific 
type of pairs, such as $(0, 1)$ pairs 
or $\big((d-1), d\big)$ pairs, respectively capturing features represented by 
minima or maxima. In the following, we will consider that each persistence 
diagram is only composed of pairs of a \emph{fixed} critical type
$\big((\criticalIndex -1), \criticalIndex)$, which will be systematically 
detailed in the experiments. 
\julienRevision{}{Note that in practice,  each critical point in the 
diagram additionally stores its 3D coordinates in $\domain$
to allow geometrical lifting 
(\autoref{sec_distance}).}

In practice, critical point pairs often directly correspond to features of 
interest in the applications and their persistence has been shown to be a 
reliable importance measure to discriminate noise from important features. In 
the diagram, the pairs located in the vicinity of the diagonal 
represent low amplitude noise while salient 
features will be associated with persistent pairs, standing out far away from 
the diagonal\EDIT{ (\autoref{fig_persistenceDiagram})}
{. For instance, in \autoref{fig_persistenceDiagram}, the two  hills are 
captured by two large pairs, while noise is encoded with smaller pairs near the 
diagonal.}
Thus, the persistence diagram has been shown in practice to be a useful, 
concise visual representation of the population of features of interest in the 
data, for which the number, range values and salience can be directly and 
visually read. 
In addition to its stability properties \cite{cohen-steiner05}, these 
observations
motivate its usage for data 
summerization, where it significantly help users distinguish salient features 
from noise.
\julienRevision{}{In the following, we review distances between persistence 
diagrams \cite{cohen-steiner05, edelsbrunner09} (\autoref{sec_distance}), 
efficient algorithms for their approximation 
\cite{Bertsekas81, Kerber2016} (\autoref{seq_auction}), as well as the 
reference 
approach for barycenter estimation 
\cite{Turner2014} (\autoref{sec_wassersteinBarycenters}).}

\subsection{Distances between Persistence diagrams}
\label{sec_distance}

To compute the barycenter of a set of persistence diagrams, a 
necessary 
ingredient is the notion of distance between them. 
Given two diagrams $\persistenceDiagram{f}$ and
$\persistenceDiagram{g}$, a pointwise distance, 
noted $\pointMetric{q}$, inspired from the $L^p$ norm, can be introduced 
in the 2D birth/death space
between 
  two points $a = (x_a, y_a) \in \persistenceDiagram{f}$ and 
$b = (x_b, y_b) \in \persistenceDiagram{g}$, with $q > 0$, as 
follows :
\vspace{-1.5ex}
\begin{equation}
\pointMetric{q}(a,b)=\left(|x_b-x_a|^q + |y_b-y_a|^q\right)^{1/q} = \|a-b\|_q
\label{eq_pointWise_metric}
\end{equation}
\vspace{-3.5ex}

\noindent
By convention, $\pointMetric{q}(a, b)$ is set to zero 
if both $a$ and $b$ exactly lie on the diagonal ($x_a = y_a$ and $x_b = y_b$).
The $q$-Wasserstein distance \cite{Kantorovich, monge81}, noted 
$\wasserstein{q}$, between $\persistenceDiagram{f}$ and 
$\persistenceDiagram{g}$ can then be introduced as:
%
%
\vspace{-1.75ex}
\begin{equation}
    \wasserstein{q}\big(\persistenceDiagram{f}, \persistenceDiagram{g}\big) = 
\min_{\phi
\in \Phi} \left(\sum_{a \in \persistenceDiagram{f}} 
\pointMetric{q}\big(a,\phi(a)\big)^q\right)^{1/q}
\label{eq_wasserstein}
\end{equation}
\vspace{-3ex}

\noindent
where $\Phi$ is the set of all possible assignments $\phi$ mapping each 
point
$a \in \persistenceDiagram{f}$ to 
a point
$b 
\in \persistenceDiagram{g}$, 
or to 
\EDIT{the 
closest diagonal point}{its projection onto the diagonal}, 
$\projection(a) = 
(\frac{x_a+y_a}{2},\frac{x_a+y_a}{2})$, which denotes the removal of the 
corresponding feature from the assignment, with a cost $\pointMetric{q}\big(a, 
\projection(a)\big)^q$ \EDIT{}{(see additional materials for an illustration)}. 
The Wasserstein distance can be computed by 
solving an 
optimal assignment problem, for which existing 
algorithms \cite{Munkres1957, Morozov:Dionysus} however often require a balanced 
setting. To 
address this, the input diagrams $\persistenceDiagram{f}$ and 
$\persistenceDiagram{g}$ are typically \emph{augmented} into
$\balancedPersistenceDiagram(f)$ and $\balancedPersistenceDiagram(g)$, which 
are obtained by injecting the diagonal projections of  all the 
points of one diagram into the other:
\vspace{-1ex}
\begin{eqnarray}
 \balancedPersistenceDiagram(f) = \persistenceDiagram{f} \cup \{ 
\projection(b) ~ | ~ b \in \persistenceDiagram{g}\}\\
  \balancedPersistenceDiagram(g) = \persistenceDiagram{g} \cup \{ 
\projection(a) ~ | ~ a \in \persistenceDiagram{f}\}
\end{eqnarray}
\vspace{-4ex}

\noindent
In this way, the Wasserstein distance is guaranteed to be preserved by 
construction, $\wasserstein{q}\big(\persistenceDiagram{f}, 
\persistenceDiagram{g}\big) = 
\wasserstein{q}\big(\balancedPersistenceDiagram(f), 
\balancedPersistenceDiagram(g)\big)$, while making the assignment problem 
balanced ($|\balancedPersistenceDiagram(f)| = |\balancedPersistenceDiagram(g)|$) 
and thus solvable with traditional assignment algorithms. 
\julienRevision{Note that w}{W}hen $q 
\rightarrow \infty$, $\wasserstein{q}$ becomes a worst case assignment distance 
called the \emph{Bottleneck} distance 
\cite{cohen-steiner05}\julienRevision{.}{, which is often interpreted in 
practice as less informative, however, than the Wasserstein distance. In the 
following, we will focus on $q = 2$.} 

In the applications, it can often be useful to geometrically lift the 
Wasserstein metric, by also taking into account
the geometrical layout of critical points \cite{soler2018}. Let 
$(c, c')$ be the 
critical point
pair corresponding the point 
$a 
\in \diagram(f)$. Let  $p_{a}^\criticalPointMixing \in 
\range^d$ be their linear combination with coefficient $\criticalPointMixing \in 
[0, 1]$ in $\domain$:  $p_{a}^\criticalPointMixing = \criticalPointMixing c' + 
(1 - 
\criticalPointMixing) c$. Our experiments (\autoref{sec_results})
only deal with
extrema\EDIT{}{,} and we set $\criticalPointMixing$ to $0$ for minima and $1$ 
for maxima 
(to only consider the extremum's location).
Then, 
 the
geometrically lifted pointwise distance $\liftedMetric{2}(a, b)$ can be 
given as:
\vspace{-2ex}
\begin{eqnarray}
\liftedMetric{2}(a, b) = 
  \sqrt{
  (1 - \geometricLifting) \pointMetric{2}(a, b)^2 
  + \geometricLifting ||p_a^\criticalPointMixing - 
p_b^\criticalPointMixing||_2^2}
\label{eq_liftedMetric}
\end{eqnarray}
\vspace{-4ex}

\noindent
The 
parameter
$\geometricLifting \in [0, 1]$ 
quantifies the importance \EDIT{that is }{}given to the geometry of  critical points 
and it must be tuned on a per application basis. 
\EDIT{When $\geometricLifting  = 1$, the combination of 
$\wasserstein{2}$ with $\liftedMetric{2}$ becomes the classical \emph{Earth 
Mover's distance} 
\cite{DBLP:conf/iccv/LevinaB01}
between critical points in 
$\domain$  and 
$\geometricLifting$ thus enables users to simply linearly blend these two 
 metrics.
 }{\vspace{-1.5ex}}
\subsection{Efficient distance computation by Auction}
\label{seq_auction}
This section briefly describes a fast approximation 
of Wasserstein distances by the auction algorithm \cite{Bertsekas81, 
Kerber2016}, which is central to our method.

%
%

The assignment problem involved in \autoref{eq_wasserstein} can be modeled in 
the form of a weighted bipartite graph, where 
the points $a \in \balancedPersistenceDiagram(f)$ are represented as nodes, 
connected by edges to nodes representing the points of 
$b \in \balancedPersistenceDiagram(g)$, with an edge weight given by 
$\pointMetric{2}(a, b)^2$ (\autoref{eq_pointWise_metric}). To efficiently 
estimate the optimal assignment, Bertsekas introduced the \emph{auction} 
algorithm \cite{Bertsekas81} (\autoref{fig_aucionExample}), which replicates the 
behavior of a real-life 
auction: the points of $\balancedPersistenceDiagram(f)$ are acting as 
\emph{bidders} that iteratively make offers for the purchase of 
the points of 
$\balancedPersistenceDiagram(g)$, known as the \emph{objects}.
Each bidder $a \in \balancedPersistenceDiagram(f)$ makes a benefit 
$\benefit_{a \rightarrow b} = -\pointMetric{2}(a, b)^2$ for the purchase of an 
object $b \in\balancedPersistenceDiagram(g)$, which is itself labeled with a 
price $\price_b \ge 0$, initially set 
to $0$. 
During the iterations of the auction, each bidder $a$ tries to purchase the 
object $b$ of highest \emph{value} $\bidValue_{a \rightarrow b} = \benefit_{a 
\rightarrow b} - \price_b$. The bidder $a$ is then said to be assigned to the 
object $b$. If $b$ was previously assigned, its previous owner becomes 
unassigned. At this stage, the price of $b$ is increased by $\priceDiff_a + 
\epsilon$, where $\priceDiff_a$ is the absolute difference between the two 
highest values $\bidValue_{a \rightarrow b}$ that the bidder $a$ found among the 
objects $b$, and where $\epsilon > 0$ is a constant. 
This bidding procedure is 
repeated iteratively among the bidders, until all bidders are assigned (which 
is guaranteed to occur by construction, thanks to the $\epsilon$ constant). At 
this point, \EDIT{we say}{it is said} that an \emph{auction round} has completed: a bijective, 
possibly sub-optimal, assignement $\phi$ exists between 
$\balancedPersistenceDiagram(f)$ and $\balancedPersistenceDiagram(g)$. The 
overall algorithm will repeat auction rounds, which 
progressively increases  prices under the effect of competing 
bidders. 

The constant $\epsilon$ plays a central role in the auction algorithm.
Let $\overallCost{2}\big(\balancedPersistenceDiagram(f), 
\balancedPersistenceDiagram(g)\big) = \sqrt{\sum_{a \in 
\balancedPersistenceDiagram(f)} \pointMetric{2}\big(a, \phi(a)\big)^2}$ be the 
approximation of the Wasserstein distance 
$\wasserstein{2}\big(\persistenceDiagram{f}, \persistenceDiagram{g}\big)$, 
obtained with the 
assignment $\phi$ returned by the algorithm.
Large  
values of $\epsilon$ will drastically accelerate convergence (as they imply 
fewer iterations 
for the construction
of a bijective assignment $\phi$ within one auction 
round, \autoref{fig_aucionExample}), while low values 
will improve the accuracy of $\overallCost{2}$.
This observation is a key insight at the basis of our approach. 
Bertsekas suggests a strategy called \emph{$\epsilon$-scaling},
which decreases  $\epsilon$ after each auction round. 
In particular, if:
\vspace{-1.5ex}
\begin{eqnarray}
 \overallCost{2}\big(\balancedPersistenceDiagram(f), 
\balancedPersistenceDiagram(g) \big)^2 \leq (1 + \auctionTolerance)^2 
\Big(\overallCost{2}\big(\balancedPersistenceDiagram(f), 
\balancedPersistenceDiagram(g) \big)^2 - 
\epsilon|\balancedPersistenceDiagram(f)|\Big)
\label{eq_stoppingCondition}
\end{eqnarray}
\vspace{-3ex}

\noindent
then it can be shown that \cite{Bertsekas91, Kerber2016}:
\vspace{-1.5ex}
\begin{eqnarray}
\wasserstein{2}\big(\persistenceDiagram{f}, 
\persistenceDiagram{g}\big)  \leq
 \overallCost{2}\big(\balancedPersistenceDiagram(f), 
\balancedPersistenceDiagram(g) \big) \leq
(1 + \auctionTolerance) 
\wasserstein{2}\big(\persistenceDiagram{f}, 
\persistenceDiagram{g}\big) 
\end{eqnarray}
\vspace{-3ex}

\noindent
This result is particularly important, as it enables to estimate the optimal 
assignment, and thus the Wasserstein distance, with an on-demand accuracy 
(controlled by the parameter $\auctionTolerance$) by 
using \autoref{eq_stoppingCondition} as a stopping condition for the overall 
auction
 algorithm.
For
persistence diagrams, Kerber et al. showed how the computation could be 
accelerated by using space partitioning data structures such as
kd-trees \cite{Kerber2016}. In practice, \EDIT{we set $\epsilon$ to be initially 
}{$\epsilon$ is initially set to be }equal to $1/4$ of the largest edge weight
$d_2(a, b)^2$, \EDIT{we divide it}{and is divided} by $5$ 
after each auction round, as recommended by Bertsekas 
\cite{Bertsekas81}\EDIT{, and we}{.} 
\julienRevision{We use 
\needsVerification{$\auctionTolerance = 0.01$}}
{$\auctionTolerance$ is set to $0.01$}
as suggested 
by 
Kerber 
et al. \cite{Kerber2016}.
\begin{algorithm}[b]
    \small
    \algsetup{linenosize=\tiny}
    \caption{\textcolor{black}{\footnotesize Reference algorithm for Wasserstein Barycenters
    \cite{Turner2014}.}} 
\label{alg:turner_algo}
\hspace*{\algorithmicindent} \textbf{Input} : Set of diagrams 
$\diagramSet = \{ \diagram(f_1), \diagram(f_2), \dots, \diagram(f_N)\}$

\hspace*{\algorithmicindent} \textbf{Output} : Wasserstein barycenter 
$\diagramBarycenter$

\begin{algorithmic}[1]

\STATE $\diagramBarycenter \leftarrow \diagram(f_i)\quad
\quad \quad \quad \quad \quad \quad \quad \quad \quad \quad
\quad\quad\quad\quad
$ // with $i$ 
randomly 
chosen in $[1, N]$

\WHILE{$\{\phi_1, \phi_2, \dots, \phi_N\}$ change}
\STATE // Relaxation start
\label{algo_startIteration}
\FOR{$i \in [1, N]$}
\STATE $\phi_i \leftarrow Assignment\big(\diagram(f_i), \diagramBarycenter\big) 
\quad ~
\quad\quad\quad\quad\quad\quad\quad\quad\quad\quad
$ // optimizing \autoref{eq_wasserstein} \label{algo_assignment}
\ENDFOR
\STATE $\diagramBarycenter \leftarrow Update(\phi_1, \dots, 
\phi_n)\quad\quad\quad\quad
\quad\quad
$ // arithmetic means in birth/death space 
\label{algo_update}
\STATE // Relaxation end
\label{algo_endIteration}
\ENDWHILE
\RETURN $\diagramBarycenter$
\end{algorithmic}
\end{algorithm}

\newcommand{\DLdeux}{{\diagramSpace_{L^2}}}
\subsection{Wasserstein barycenters of Persistence diagrams}
\label{sec_wassersteinBarycenters}
Let $\diagramSpace$ be the space of persistence diagrams. 
The \needsVerification{discrete} \emph{Wasserstein barycenter} 
of a set $\diagramSet = \{\persistenceDiagram{f_1}, \persistenceDiagram{f_2}, 
\dots, \persistenceDiagram{f_N}\}$ of persistence diagrams can  be 
introduced as the Fr\'echet mean of the set, under 
the metric $\wasserstein{2}$.
It is 
the diagram $\diagramBarycenter$
that 
minimizes its 
distance to all the diagrams of the set (\emph{i.e.} minimizer of the so-called
Fr\'echet energy):
\vspace{-1.5ex}
\begin{eqnarray}
\label{eq_frechet}
 \diagramBarycenter = \argmin_{\diagram \in \diagramSpace} 
\sum_{\persistenceDiagram{f_i} \in \diagramSet} \wasserstein{2}\big(\diagram, 
\persistenceDiagram{f_i}\big)^2
\end{eqnarray}
\vspace{-3ex}

The computation of Wasserstein barycenters  involves a computationally 
demanding optimization problem, for which the existence of at least one 
locally optimum solution has been shown by Turner et al. \cite{Turner2014}, 
who also introduced 
the first algorithm for its computation. This algorithm 
(\autoref{alg:turner_algo}) 
consists in iterating a procedure that we call \emph{Relaxation} (line 
\ref{algo_startIteration} to \ref{algo_endIteration}), 
which resembles a Lloyd relaxation \cite{lloyd82}, and
which is composed itself 
of two sub-routines: \emph{(i) Assignment} (line \ref{algo_assignment}) and 
\emph{(ii) Update} (line \ref{algo_update}).
Given an initial 
barycenter candidate $\diagram$ randomly chosen among the set $\diagramSet$, the 
first step (\emph{(i) Assignment}) consists in computing an optimal 
assignment $\phi_i : \diagram 
\rightarrow \diagram(f_i)$ between $\diagram$ and each diagram $\diagram(f_i)$ 
of the set $\diagramSet$, with regard to \autoref{eq_wasserstein}. The second 
step (\emph{(ii) Update}) consists in 
updating the candidate $\diagram$ to a position in $\diagramSpace$ which 
minimizes the sum of
its squared distances 
to the diagrams of $\diagramSet$ under 
the current set of 
assignments $\{\phi_1, \phi_2, \dots, \phi_N\}$. In practice, this last step is 
achieved by \EDIT{simply}{} replacing 
each point $a \in \diagram$ by the arithmetic mean (in the birth/death space) of 
all its assignments $\phi_i(a)$. The overall algorithm continues to iterate 
the \emph{Relaxation} procedure
until the set of optimal assignments 
$\phi_i$
remains identical for two consecutive 
iterations.

\EDIT{
    \subsection{Overview of our approach}
    \label{overview}
}{}
The reference 
algorithm for Wasserstein barycenters (\autoref{alg:turner_algo}) reveals 
impractical for real-life data sets. Its main computational bottleneck is the 
\emph{Assignment} step, which involves the computation of $N$ Wasserstein 
distances (\autoref{eq_wasserstein}). However, as detailed in the result 
section (\autoref{sec_results}), even when combined with 
efficient algorithms for the Wasserstein distance exact computation 
\cite{soler2018} or even approximation \cite{Kerber2016} 
(\autoref{seq_auction}), this overall method can still lead to hours of 
computation in practice.

Thus, a drastically different approach is needed to improve computation 
efficiency, especially for applications such as ensemble clustering, which 
require multiple barycenter estimations per iteration.

\EDIT{
In this work, we introduce a novel progressive framework for Wasserstein 
barycenters, which is based on two key observations. First, from one 
\emph{Relaxation}
iteration 
to the next (lines \autoref{algo_startIteration} to 
\autoref{algo_endIteration}, \autoref{alg:turner_algo}), the estimated 
barycenter is likely to  vary only slightly. This motivates the design of a 
faster algorithm 
for the \emph{Assignment} step,
which would be capable 
of starting its 
optimization from a relevant initial assignment, typically given in our setting 
by the previous \emph{Relaxation} iteration. 
Second, 
in the initial \emph{Relaxation}
iterations, the 
estimated barycenter can be arbitrarily far from the final, optimized 
barycenter. Thus, for these iterations, it can be  beneficial to 
relax the level of accuracy of the \emph{Assignment} step, 
which is the main  bottleneck, and to progressively increase it 
when it is the most needed, as the barycenter converges to a solution.

%
%
%
%

\section{Progressive Barycenters}
This section presents our novel progressive framework for the approximation of 
Wasserstein barycenters of a set of Persistence diagrams $\diagramSet = 
\{\diagram(f_1), \diagram(f_2), \dots, \diagram(f_N)\}$. As mentioned above,
the key insights of our approach are twofolds. First, the assignments involved 
in the Wasserstein distance estimations can be re-used as initial 
conditions along the iterations of the barycenter \emph{Relaxation} 
(\autoref{sec_assignmentReuse}). Second, progressivity can be injected in the 
process to accelerate convergence, by 
adequately controlling the level of accuracy in the evaluation of the 
Wasserstein distances along the \emph{Relaxation} iterations of the barycenter. 
This progressivity can be injected at two 
levels: by controlling the accuracy of the distance estimation 
itself (\autoref{sec_partialbidding}) and the resolution of the input diagrams
(\autoref{sec_progressivePersistence}). The rest of the
section 
presents a parallelization of our framework (\autoref{sec_parallelism}) and
describes 
an interruptible algorithm, capable of 
respecting running time constraints (\autoref{sec_interruptible}).
}
{
    \vspace{-2ex}
\section{Progressive Barycenters}
This section presents our novel progressive framework for the approximation of
Wasserstein barycenters of a set of Persistence 
diagrams\julienRevision{$\diagramSet =
\{\diagram(f_1), \diagram(f_2), \dots, \diagram(f_N)\}$.}{.}
\subsection{Overview}
The key insights of
our approach are twofolds. First, in the reference algorithm 
(\autoref{alg:turner_algo}), from one 
\emph{Relaxation}
iteration 
to the next (lines \autoref{algo_startIteration} to 
\autoref{algo_endIteration}), the estimated 
barycenter is likely to  vary only slightly. 
\julienRevision{Previous assignments can thus}
{Thus, the assignments involved in the Wasserstein distance estimations can}
be re-used as initial conditions 
along the iterations
of the barycenter \emph{Relaxation}
(\autoref{sec_assignmentReuse}). 
Second, 
in the initial \emph{Relaxation}
iterations, the 
estimated barycenter can be arbitrarily far from the final, optimized 
barycenter. Thus, for these early iterations, it can be  beneficial to 
relax the level of accuracy of the \emph{Assignment} step, 
\EDIT{which is the main  bottleneck,} and to progressively increase it 
as the barycenter converges to a solution. Progressivity can be injected at two levels: by controlling the accuracy
of the distance estimation itself (\autoref{sec_partialbidding}) and the
resolution of the input diagrams (\autoref{sec_progressivePersistence}). Our framework is
easily parallelizable
(\autoref{sec_parallelism}) and the progressivity allows to design an interruptible algorithm, capable
of respecting running time constraints (\autoref{sec_interruptible}).
}

\begin{figure*}
\includegraphics[width=\textwidth]
{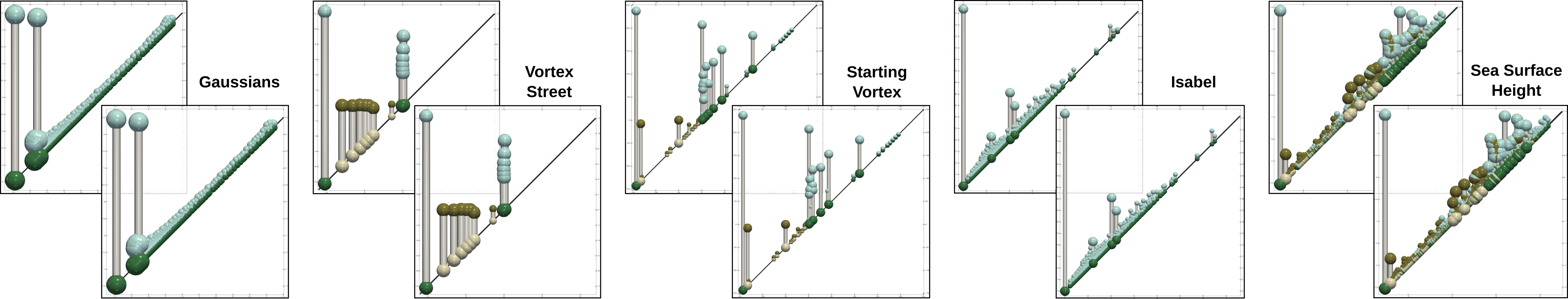}
\vspace{-1ex}
\imageCaption{
Visual comparison between the converged Wasserstein barycenters obtained with 
the \emph{Auction barycenter} algorithm 
\mixRef
(top) and 
our approach (bottom) for one cluster 
(\autoref{sec_clusteringResults}) 
of 
each  ensemble data set.
Differences are barely noticeable, and only for small persistence pairs. 
\finalVersionSpace{-1ex}
}
\label{fig_visualComparison}
\end{figure*}

\subsection{Auctions with Price Memorization}
\label{sec_assignmentReuse}
The \emph{Assignment} step of the Wasserstein barycenter computation (line 
\ref{algo_assignment}, \autoref{alg:turner_algo}) can be resolved in principle 
with any of the existing techniques for Wasserstein distance 
estimation \cite{Munkres1957,  Morozov:Dionysus, Kerber2016, soler2018}. 
Among them, the Auction based approach \cite{Bertsekas81, 
Kerber2016} (\autoref{seq_auction}) is particularly relevant as it can compute 
very efficiently approximations with on-demand accuracy. 

In the following,
we consider that each distance computation involves \emph{augmented} diagrams 
\EDIT{(\autoref{sec_distance})}{}.
Each input diagram 
$\diagram(f_i) \in \diagramSet$ is then considered as a set of
\emph{bidders} \EDIT{(\autoref{seq_auction})}{} while the output barycenter 
$\diagramBarycenter$ contains the \emph{objects} to purchase. 
Each input diagram $\diagram(f_i)$ maintains its own list of prices 
$\price_b^i$ for the purchase of the objects $b \in \diagramBarycenter$ by the 
bidders $a \in \diagram(f_i)$. 
The search by a bidder for the two most valuable objects to purchase 
is
accelerated with space partitioning data structures,
by using a kd-tree and a lazy heap respectively for the 
off- and on-diagonal points \cite{Kerber2016} (these structures are 
re-computed for  each \emph{Relaxation}). 
\needsVerification{Thus, the output barycenter 
$\diagramBarycenter$ 
maintains only one kd-tree and one lazy heap for this purpose}. Since 
Wasserstein distances are only approximated in this strategy, we suggest to 
relax the overall stopping condition (\autoref{alg:turner_algo}) and stop the 
iterations after two successive increases in Fr\'echet energy 
(\autoref{eq_frechet})\needsVerification{, as commonly done in gradient 
descent optimizations}. 
\needsVerification{In the rest of the paper, we call the above strategy the 
\emph{Auction barycenter algorithm \mixRef},
as it \EDIT{naively}{just} combines the algorithms 
by Turner et al. \cite{Turner2014} and Kerber et al. \cite{Kerber2016}.}

%
However, 
this \EDIT{naive}{} usage of the auction algorithm 
results in a complete reboot of the entire sequence of auction rounds upon each 
\emph{Relaxation}, while in practice, for the barycenter problem, the 
output assignments $\phi_i$ may be very similar from one 
\emph{Relaxation} iteration to the next 
and thus could be re-used as initial solutions. For this, we 
introduce 
a mechanism that we call \emph{Price Memorization}, which \EDIT{simply}{} consists in 
initializing the 
prices $\price_b^i$  for each bidder $a \in \diagram(f_i)$ to the 
prices obtained at the previous \emph{Relaxation} iteration (instead of $0$). 
This has the positive effect of encouraging bidders to bid in priority on 
objects which were previously assigned to them, hence effectively re-using the 
previous assignments as an initial solution. This memorization makes most of 
the 
early auction rounds become unnecessary in practice, 
which enables  to 
drastically reduce their number, as detailed in the following.

\subsection{Accuracy-driven progressivity}
\label{sec_partialbidding}

The reference algorithm for Wasserstein barycenter computation 
(\autoref{alg:turner_algo}) can also be interpreted as a variant of gradient 
descent \cite{Turner2014}. For such methods, it is often observed 
that approximations of the gradient, instead of exact computations, can be 
sufficient in practice to reach convergence. This observation is at the basis 
of our progressive strategy. Indeed, in the early \emph{Relaxation} iterations, 
the  barycenter can be arbitrarily far from the converged result and 
achieving a high accuracy in the \emph{Assignment} step (line 
\ref{algo_assignment}) for these iterations is often 
a waste of computation time. Therefore we introduce a mechanism that 
progressively increases the accuracy of the \emph{Assignment} step along the 
\emph{Relaxation} iterations, 
\EDIT{to provide the maximum}{in order to obtain more} 
accuracy near convergence.

To achieve this, inspired by the internals of the auction algorithm, we apply a 
\emph{global} $\epsilon$-scaling \EDIT{(\autoref{seq_auction})}{}, 
where we 
progressively decrease the value of $\epsilon$, but only at the end of each 
\emph{Relaxation}. Combined with \emph{Price Memorization} 
\EDIT{(\autoref{sec_assignmentReuse})}{}, this strategy enables us to perform \emph{only 
one} 
auction round per \emph{Assignment} step.
As large 
$\epsilon$ values accelerate auction convergence at the price of accuracy, this 
strategy effectively 
speeds up
the early \emph{Relaxation} iterations 
and leads to more and more accurate auctions, and thus assignments, along the
\emph{Relaxation} iterations.

In practice, we divide $\epsilon$ by $5$ after each \emph{Relaxation}, 
as suggested by Bertsekas \cite{Bertsekas81} in the case of the 
regular auction algorithm (\autoref{seq_auction}). Moreover, to guarantee 
precise final barycenters (obtained for small $\epsilon$ values), we modify the 
overall stopping condition to prevent the
algorithm
from stopping 
if $\epsilon$ is \needsVerification{larger than $10^{-5}$ of its initial 
value.} 

\begin{table}
\finalVersionSpace{-1ex}
\caption{Comparison of running times (in seconds, 1 thread)
for the 
estimation of Wasserstein barycenters of Persistence diagrams. 
$N$ and $\#_{\diagram(f_i)}$ respectively stand for 
the number of members in the 
ensemble and the average size of the input persistence diagrams.
}
    \centering
    \renewcommand{\citepunct}{+}
    \resizebox{\columnwidth}{!}{
\begin{tabular}{|l|rr||r|rrr|r|}
    \hline
    \vspace{1pt}
    Data set            
    & $N$   & $\#_{\diagram(f_i)}$      
    & \pbox{10cm}{ Sinkhorn\\ \centering\cite{lacombe2018}} 
    & \pbox{10cm}{ 
Munkres\\\centering\cite{Turner2014}+\hspace{-.05ex}\cite{soler2018}}   & 
\pbox{10cm}{Auction\\\centering\mixRef}
  &  \textbf{Ours}  & Speedup     \\
    \hline
Gaussians (\autoref{fig_gaussians})
& 100   
& 2,078    
  &7,499.33
  &$>24$H& 8,975.60 
   &  785.53&11.4\\
Vortex Street (\autoref{fig_vortexStreet})
& 45    
& 14    
  &54.21
  &0.14       &0.47       
      &  0.23&0.6     \\
Starting Vortex (\autoref{fig_startingVortex})
& 12    
& 36    
  &40.98
  &0.06       &0.67       
        &  0.28&0.2     \\
Isabel (3D) (\autoref{fig:teaser})
& 12   
& 1,337
  &1,070.57
  &\needsVerification{$>24$H}      & 377.42    
     &   82.95&4.5    \\
\needsVerification{Sea Surface Height (\autoref{fig_seaSurfaceHeight})}
& 48    
&1,379 
  & 4,565.37 
  &$>24$H   &949.08 
     & 75.90
     &12.5  \\
    \hline
\end{tabular}
}
\label{table_timings}
\finalVersionSpace{-1ex}
\end{table}

\subsection{Persistence-driven progressivity} 
\label{sec_progressivePersistence}

In practice, the persistence diagrams of real-life data sets  
often contain 
a very large number of critical point pairs of low persistence. These numerous 
small pairs correspond to noise and are often meaningless for the 
applications. However, although they individually have only little impact on 
Wasserstein distances (\autoref{eq_wasserstein}), their overall contributions 
may be non-negligible. To account for this, we introduce in this 
section a persistence-driven progressive mechanism, which progressively inserts 
in the input diagrams critical point pairs of decreasing persistence. This 
focuses the early \emph{Relaxation} iterations on the most salient features of 
the data, while considering the noisy ones last. In practice,
this encourages the optimization to explore more relevant local minima of the 
Fr\'echet energy (\autoref{eq_frechet}) that favor persistent features.

Given an input diagram $\diagram(f_i)$, let 
$\diagram_\persistenceThreshold(f_i)$ be the subset of its points with 
persistence higher than $\persistenceThreshold$: 
$\diagram_\persistenceThreshold(f_i) = \{a \in \diagram(f_i) ~| ~
y_a - x_a > \persistenceThreshold\}$. To account for persistence-driven 
progressivity, we run our barycenter algorithm (with \emph{Price Memorization}, 
\autoref{sec_assignmentReuse}, and accuracy-driven progressivity, 
\autoref{sec_partialbidding}) by initially considering as an input the diagrams 
$\diagram_\persistenceThreshold(f_i)$.
After each \emph{Relaxation} iteration (\autoref{alg_final}, line 
\ref{line_persistenceScaling}),
we decrease 
$\persistenceThreshold$
such that $|\diagram_\persistenceThreshold(f_i)|$ does not 
increase 
by more than 10\%  (to progress at uniform 
speed)
and such that 
$\persistenceThreshold$ does not get smaller than  
$\sqrt{\persistenceScaling\epsilon}$ 
(we set $\persistenceScaling = 4$ to replicate locally Bertsekas's suggestion 
for $\epsilon$ setting, \autoref{seq_auction}).
Initially, 
$\persistenceThreshold$ is set to half of the maximum persistence found in the 
input diagrams. Along the \emph{Relaxation} iterations, the input 
diagrams $\diagram_\persistenceThreshold(f_i)$ are progressively populated, 
which yields the 
introduction of new points in the barycenter $\diagramBarycenter$, which we 
initialize at locations selected 
uniformly 
among the newly introduced points of the $N$ inputs. This strategy enables to 
\emph{distribute} among the inputs the initialization of the new barycenter 
points.
The corresponding prices 
%
are initialized with the minimum price $\price_b^i$ found for the objects $b 
\in \diagramBarycenter$ at the previous iteration. 

\begin{algorithm}[b]
    \small
    \algsetup{linenosize=\tiny}
    \caption{\footnotesize Our overall algorithm for Progressive Wasserstein 
Barycenters.}
\label{alg_final}
\hspace*{\algorithmicindent} \textbf{Input} : Set of diagrams 
$\diagramSet = \{ \diagram(f_1), \diagram(f_2), \dots, \diagram(f_N)\}$, time 
constraint $\timeConstraint$

\hspace*{\algorithmicindent} \textbf{Output} : Wasserstein barycenter 
$\diagramBarycenter_\persistenceThreshold$

\begin{algorithmic}[1]

\STATE $\diagramBarycenter_\persistenceThreshold \leftarrow 
\diagram_\persistenceThreshold(f_i)\quad
\quad \quad \quad \quad \quad \quad \quad \quad \quad \quad
\quad \quad \quad
\quad 
$ // with $i$ 
randomly 
chosen in $[1, N]$

\WHILE{the Fr\'echet energy decreases}
\STATE // Relaxation start
\label{algo_final_startIteration}
\FOR{$i \in [1, N]$}
\STATE // In parallel 
$
\quad \quad \quad \quad
\quad \quad \quad
\quad \quad \quad
\quad \quad \quad
\quad \quad \quad \quad \quad \quad\quad \quad 
~ ~ 
$\textbf{// \autoref{sec_parallelism}}
\STATE $\phi_i \leftarrow 
Assignment\big(\diagram_\persistenceThreshold(f_i), 
\diagramBarycenter_\persistenceThreshold\big) 
\quad \quad ~
\quad \quad \quad \quad \quad \quad\quad \quad \quad \quad \quad ~ ~  ~
$ \textbf{// \autoref{sec_assignmentReuse}}
\ENDFOR
\STATE $\diagramBarycenter_\persistenceThreshold \leftarrow Update(\phi_1, 
\dots, 
\phi_n)\quad\quad\quad\quad \quad \quad$ // arithmetic means in birth/death 
space 
\label{algo_final_update}
\STATE $EpsilonScaling() \quad \quad \quad
\quad \quad \quad
\quad \quad \quad
\quad \quad \quad
\quad \quad \quad \quad \quad \quad\quad \quad ~ ~ ~
$ \textbf{// \autoref{sec_partialbidding}}
\STATE \textbf{if} $t < 0.1 \times \timeConstraint$ \textbf{then} 
$PersistenceScaling() \quad \quad ~ ~ ~
\quad \quad \quad \quad \quad \quad\quad \quad $
\textbf{// \autoref{sec_progressivePersistence}}
\label{line_persistenceScaling}
\STATE \textbf{else if} $t >= \timeConstraint$ \textbf{then return} 
$\diagramBarycenter_\persistenceThreshold \quad\quad\quad\quad
\quad\quad ~ ~
\quad \quad \quad \quad \quad \quad\quad \quad 
$ \textbf{// \autoref{sec_interruptible}}
%
\STATE // Relaxation end
\label{algo_final_endIteration}
\ENDWHILE
\RETURN $\diagramBarycenter_\persistenceThreshold$
\end{algorithmic}
\end{algorithm}

\subsection{Parallelism}
\label{sec_parallelism}
Our progressive framework can be trivially parallelized as the most
computationally demanding task, the \emph{Assignment} step
(\autoref{alg_final}), is independent for each input diagram 
$\diagram(f_i)$. The space partitioning data structures used 
for
proximity queries to $\diagramBarycenter$ are accessed independently by each 
bidder diagram. 
Thus, we 
parallelize our approach by running the \emph{Assignment} step in 
$\threadNumber$ independent threads.




\subsection{Computation time constraints}
\label{sec_interruptible}

Our persistence-driven progressivity (\autoref{sec_progressivePersistence})
focuses the early iterations of the optimization on the most salient features, 
while considering the noisy ones last. However, as discussed before, low 
persistence pairs in the input diagrams are often considered as meaningless in 
the applications. This means that our progressive framework can in principle
 be \emph{interrupted} before convergence and still provide a meaningful 
result. 

Let $\timeConstraint$ be a user defined time constraint. We first progressively 
introduce points in the input diagrams $\diagram_\persistenceThreshold(f_i)$ and 
perform the 
\emph{Relaxation} iterations for the first $10\%$ of the 
time constraint 
$\timeConstraint$, as described in 
\autoref{sec_progressivePersistence}. At this point, the optimized barycenter 
$\diagramBarycenter$ contains only a fraction of the points it would have 
contained if computed until convergence. To guarantee a precise output 
barycenter, we found that continuing the \emph{Relaxation} iterations for the 
remaining $90\%$ of the time, without introducing new persistence pairs, provided 
the best results. \julienRevision{Note that i}{I}n practice, in most of our 
experiments, we observed
that this second optimization part fully converged even before reaching 
$90\%$ of the computation time constraint. \autoref{alg_final} summarizes our 
overall approach for Wasserstein barycenters of persistence diagrams, 
with 
price memorization, 
progressivity, parallelism and 
time constraints. \julienRevision{For clarity, we provide in the additional 
materials a toy 
example of the unfolding of our algorithm.}
{Several iterations of our algorithm are illustrated 
on a toy example in additional material.}

\section{Application to Ensemble Topological Clustering}
\label{sec_clustering}
This section presents an application of our progressive framework for 
Wasserstein barycenters of persistence diagrams to the clustering of the 
members of ensemble data sets. Since it focuses on persistence diagrams, this 
strategy enables to group together ensemble members that have 
the same topological profile, 
hence effectively highlighting the main trends found in 
the ensemble in terms of features of interest.


The $k$-means is a popular algorithm for the clustering of the elements of a 
set, if distances and barycenters can be estimated on this set. The latter 
is efficiently computable for persistence diagrams thanks to our novel 
progressive framework and we detail in the following how to revisit the 
$k$-means algorithm to make it use our progressive barycenters \EDIT{}{ as estimates
of the centroids of the clusters}.

\begin{table}
\caption{Running times 
(in seconds) of our approach (run until convergence) with 1 and 8 threads.
$N$ and $\#_{\diagram(f_i)}$ 
stand for 
the number of members in the 
ensemble and the average size of the 
diagrams.
}
\label{table_parallel}
    \centering
    \renewcommand{\citepunct}{+}
    \resizebox{\columnwidth}{!}{
\begin{tabular}{|l|rr||rr|r|}
    \hline
    \vspace{1pt}
    Data set       & $N$ & $\#_{\diagram(f_i)}$   
    & 1 thread   & 8 threads & Speedup   
       \\
    \hline
    Gaussians  (\autoref{fig_gaussians}) &100 & 2,078&           785.53 &  
117.91  & 6.6\\
    Vortex Street  (\autoref{fig_vortexStreet}) & 45 & 14          & 0.23    & 
0.10 & 2.3\\
    Starting Vortex  (\autoref{fig_startingVortex}) & 12 & 36          &  0.28  
 
   & 0.19 & 1.5      \\
    Isabel (3D)  (\autoref{fig:teaser})   & 12 & 1,337           & 82.95    & 
  31.75 & 2.6\\
    \needsVerification{Sea Surface Height  (\autoref{fig_seaSurfaceHeight})} 
& 48 & 1,379
&75.90
 
 & 19.40
 & 3.9\\
    \hline
\end{tabular}
}
\end{table}

\begin{table}
\caption{Comparison of Fr\'echet energy (\autoref{eq_frechet}) 
at convergence
between 
 the \emph{Auction barycenter} method 
 \mixRef
\hspace{.25ex} and our 
approach. 
}
\label{table_energy}
\centering
\resizebox{\columnwidth}{!}{
\begin{tabular}{|l|rr||rr|r|}
    \hline
    \vspace{1pt}
    Data set & $N$ & $\#_{\diagram(f_i)}$ & 
\pbox{10cm}{Auction\\\centering\mixRef}
& Ours & 
\textbf{Ratio}\\
    \hline

    Gaussians  (\autoref{fig_gaussians}) & 100 & 2,078 &39.4 &39.0  & 0.99\\
    Vortex Street (\autoref{fig_vortexStreet})   & 12         &  36  
                                                 & 415.1
&412.5
& 0.99\\
    Starting Vortex (\autoref{fig_startingVortex}) & 45         & 14 
&  112,787.0
&112,642.0
& 1.00\\
    Isabel (3D)  (\autoref{fig:teaser}) & 12            &  1,337  
& 2,395.6 
& 2,337.1 
& 0.98\\ 
    Sea Surface Height (\autoref{fig_seaSurfaceHeight})
& 48  & 1,379
&  7.2
& 7.1 
& 0.99\\
    

    \hline
\end{tabular}
}
\end{table}

\begin{figure}[t]
    \includegraphics[width=\linewidth]{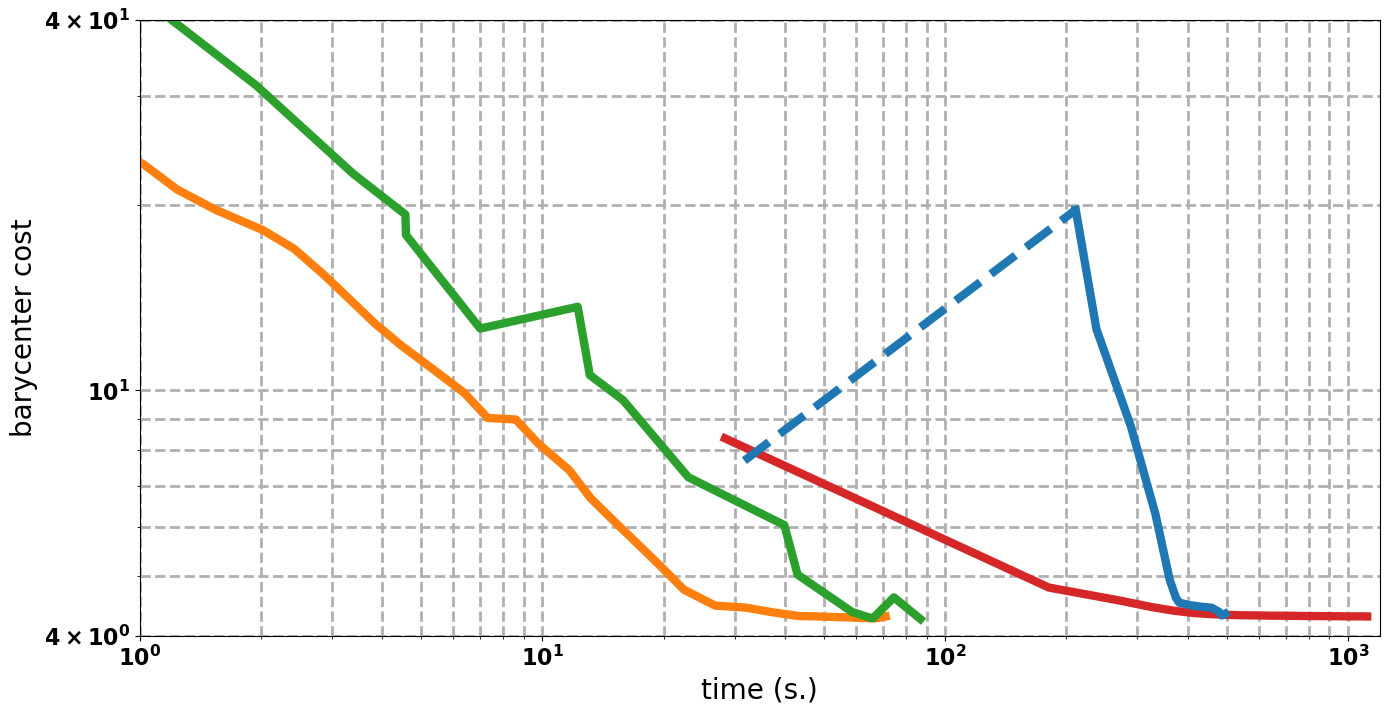}
    \finalVersionSpace{-2ex}
    \imageCaption{Comparison of the 
    evolution of the
    Fr\'echet energy 
    (log scale, \emph{Sea 
Surface Height}, maximum diagrams), for the 
\emph{Auction barycenter} method 
\mixRef
(red) 
and 3 
variants of our approach: without (blue) and with (orange) persistence 
progressivity, and with 
time constraints (green). 
Persistence-driven progressivity drastically accelerates convergence.
}
\finalVersionSpace{-1ex}
\label{fig_convergence}
%
\end{figure}

The $k$-means is an iterative algorithm, which highly resembles barycenter 
computation algorithms (\autoref{sec_wassersteinBarycenters}),
where each \emph{Clustering} iteration 
is composed itself of two sub-routines: \emph{(i) Assignment} and \emph{(ii) 
Update}. Initially, $k$ cluster centroids $\diagramBarycenter_j$ ($j \in [1, 
k]$) are initialized on $k$ diagrams $\diagram(f_i)$ of the input set 
$\diagramSet$. For this, in practice, we use the \emph{$k$-means++} heuristic 
described by Celebi et al. \cite{celebi13}, which aims at maximizing the 
distance between centroids. Then, the \emph{Assignment} step consists in 
assigning each diagram $\diagram(f_i)$ to its closest centroid 
$\diagramBarycenter_j$. This implies the computation, for each diagram 
$\diagram(f_i)$ of its Wasserstein distance $\wasserstein{2}$ to all the 
centroids $\diagramBarycenter_j, j \in [1, k]$. 
For this step, we estimate 
these pairwise distances with the Auction algorithm run 
until convergence ($\auctionTolerance = 0.01$, \autoref{seq_auction}).
In practice, we use the \emph{accelerated 
$k$-means} strategy described by Elkan \cite{elkan03}, which exploits the 
triangle inequality between centroids to skip, given a diagram $\diagram(f_i)$, 
the computation of its distance to centroids $\diagramBarycenter_j$ which 
cannot be the closest. Next, the \emph{Update} step consists in updating each 
centroid's location by placing it at the barycenter (with \autoref{alg_final}) 
of its assigned diagrams 
$\diagram(f_i)$. The algorithm continues these \emph{Clustering} iterations 
until convergence, \emph{i.e.} until the assignments between the diagrams 
$\diagram(f_i)$ and the $k$ centroids $\diagramBarycenter_j$ do not evolve 
anymore, hence yielding the final clustering.

%
From our experience, the \emph{Update} step of a \emph{Clustering} 
iteration is by far the most computationally expensive. To speed up this stage 
in practice, we derive a strategy that is similar to our approach for 
barycenter approximation: 
we reduce the computation load of each 
\emph{Clustering} iteration and progressively increase their accuracy along the 
optimization. This strategy is motivated by a similar observation: early 
centroids are quite different from the converged ones, which motivates 
an accuracy reduction in the early \emph{Clustering} iterations of the 
algorithm. 
Thus, 
for each \emph{Clustering} iteration,
we use a single 
round of auction with price memorization
(\autoref{sec_assignmentReuse}), and a 
\emph{single} barycenter update  (\emph{i.e.} a single 
\emph{Relaxation} iteration, 
\autoref{alg_final}). Overall, only one 
global $\epsilon$-scaling (\autoref{sec_partialbidding}) is applied at the end 
of each \emph{Clustering} iteration. This enhances the $k$-means algorithm with 
accuracy progressivity. 
If a diagram $\diagram(f_i)$ migrates from a cluster $j$ to a cluster $l$, the 
prices of the objects of $\diagramBarycenter_l$ for the bidders of 
$\diagram(f_i)$ are initialized to $0$ and we run the 
auction algorithm  between 
$\diagram(f_i)$ and $\diagramBarycenter_l$ until 
\needsVerification{the pairwise $\epsilon$ value 
matches the global $\epsilon$ value}, in order to obtain prices for 
$\diagram(f_i)$  which are comparable to the other diagrams.
Also, we apply persistence-driven progressivity 
(\autoref{sec_progressivePersistence}) by 
adding 
persistence pairs of decreasing persistence in each diagram $\diagram(f_i)$ 
along the \emph{Clustering} iterations. Finally, a computation time constraint 
can also be provided, 
\needsVerification{as described in \autoref{sec_interruptible}}. Results of our 
clustering scheme are presented in \autoref{sec_clusteringResults}.

\section{Results}
\label{sec_results}
This section presents experimental results obtained on a ècomputer with 
two Xeon CPUs (3.0 GHz, 2x4 cores), with 64GB of RAM. 
The  input persistence diagrams were computed with 
\julienRevision{the algorithm by Gueunet et al. 
\cite{gueunet_ldav17}, which is available 
in the Topology ToolKit (TTK) \cite{ttk17}.}
{the FTM algorithm \cite{gueunet_ldav17, ttk17}.}
We implemented our approach in C++, 
as TTK modules.

\begin{figure}
\includegraphics[width=\columnwidth]{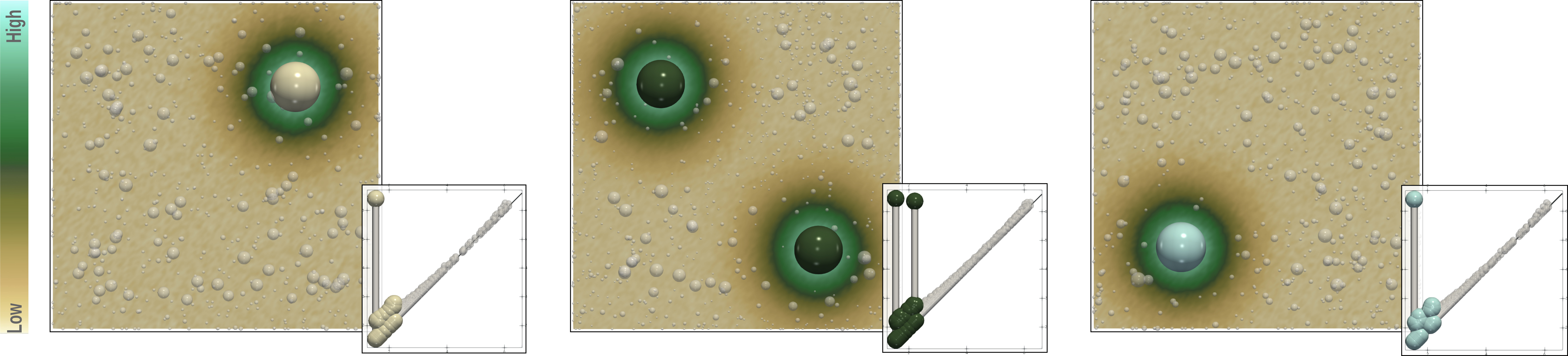}
 \imageCaption{Clustering the \emph{Gaussians} ensemble. From left to 
right, pointwise mean and Wasserstein barycenter for each of the identified 
clusters ($\timeConstraint: 10 s.$)
with geometrical lifting \needsVerification{
($\geometricLifting = 0.65$)}.}
\label{fig_gaussians}
\end{figure}

Our experiments were performed on a variety of simulated and acquired 2D and 3D 
ensembles,
\julienRevision{\EDIT{.}{, exactly the same as those used by Favelier et al. 
\cite{favelier2018}.}}
{taken from Favelier et al. \cite{favelier2018}.}
The \emph{Gaussians} ensemble contains 100 2D synthetic 
noisy members, with 
3 patterns of Gaussians (\autoref{fig_gaussians}). The 
\julienRevision{}{considered} features 
of interest in this example are the maxima\julienRevision{, hence 
only the diagrams including the $\big((d- 1), d\big)$ persistence pairs are 
considered.}{.}
The \emph{Vortex Street} ensemble (\autoref{fig_vortexStreet}) includes 45 runs 
of a 2D simulation of flow turbulence behind an obstacle. The considered scalar 
field is the curl orthogonal component, 
for 5 fluids of different viscosity\julienRevision{(9 runs per fluid, each run 
with varying Reynolds 
numbers).}{.} In this application, salient extrema are typically considered as 
reliable estimations of the center of vortices.
Thus, each run is represented by two diagrams, processed 
independently by our algorithms: one for the $(0, 1)$ pairs (involving minima) 
and one for the $\big((d- 1), d\big)$ pairs (involving maxima).
The \emph{Starting Vortex} ensemble (\autoref{fig_startingVortex}) includes 12 
runs of a 2D simulation of the formation of a vortex behind a wing, for 2 
distinct wing configurations. The considered data is also the curl 
orthogonal component and diagrams involving minima and maxima are also 
considered. The \emph{Isabel} data set (\autoref{fig:teaser}) is 
a volume ensemble of 12 members, showing key time 
steps (formation, 
drift and landfall) in the simulation of the Isabel hurricane 
\cite{scivisIsabel}. In this example, the eyewall of the hurricane is typically 
characterized by high wind velocities, well captured by velocity maxima. Thus 
we only consider diagrams involving maxima.
Finally, the \emph{Sea Surface Height} ensemble 
(\autoref{fig_seaSurfaceHeight}) is composed of 
48 observations taken in January, April, July and October 2012 
(\href{https://ecco.jpl.nasa.gov/products/all/}{https://ecco.jpl.nasa.gov/products/all/}). 
\julienRevision{The}{Here, the}
features of 
interest 
are the center of  eddies, which can be 
reliably estimated with 
height extrema. Thus, both the diagrams involving the minima and maxima are 
considered and 
independently processed by our algorithms.
Unless stated otherwise, all 
results were obtained by considering the Wasserstein metric $\wasserstein{2}$ 
based on the original pointwise metric (\autoref{eq_pointWise_metric}) without 
geometrical lifting (\emph{i.e.} $\geometricLifting = 0$, 
\autoref{sec_distance}).

%
%
%
%
%
%
%

\subsection{Time performance}
\label{sec_time_perf}
\julienRevision{We first evaluate the time 
performance of our progressive framework 
when run until convergence (\emph{i.e.} no computation time constraint). 
The corresponding timings are reported in 
\autoref{table_timings}.}
{\autoref{table_timings} evaluates the time 
performance of our progressive framework 
when run until convergence (\emph{i.e.} no computation time constraint).}
This table also provides running times for 3 
alternatives. 
The column, 
\emph{Sinkhorn}, provides the timings obtained with a 
\needsVerification{Python} CPU implementation
kindly provided by 
Lacombe et al. \cite{lacombe2018}, 
for which we used 
the recommended parameter values 
(entropic term: 
$10^{-1}/\#_{\diagram(f_i)}$ 
heat 
map resolution: $100^2$). 
Note that this approach casts the 
problem as an Eulerian transport optimization under an entropic 
regularization term.
Thus, it optimizes for a convex functional which is considerably different from 
the Fr\'echet energy considered in our approach (\autoref{eq_frechet}). 
Overall, these aspects, in addition to the difference in programming language,
challenge 
direct comparisons and we only report running times for completeness.
The columns \emph{Munkres}, noted 
\cite{Turner2014}+\hspace{-.05ex}\cite{soler2018}, and 
\emph{Auction}, noted 
\mixRef,
report the 
running times of our own C++ implementation of Turner's algorithm 
\cite{Turner2014} where  
distances 
are respectively estimated with the 
exact method by Soler et al. \cite{soler2018}
\julienRevision{(available in TTK \cite{ttk17})}{}
and our own C++ implementation of the auction-based approximation by Kerber et 
al. \cite{Kerber2016} (with kd-tree and lazy heap, run until convergence,
\needsVerification{$\auctionTolerance = 0.01$}).

\begin{figure}
\includegraphics[width=\columnwidth]{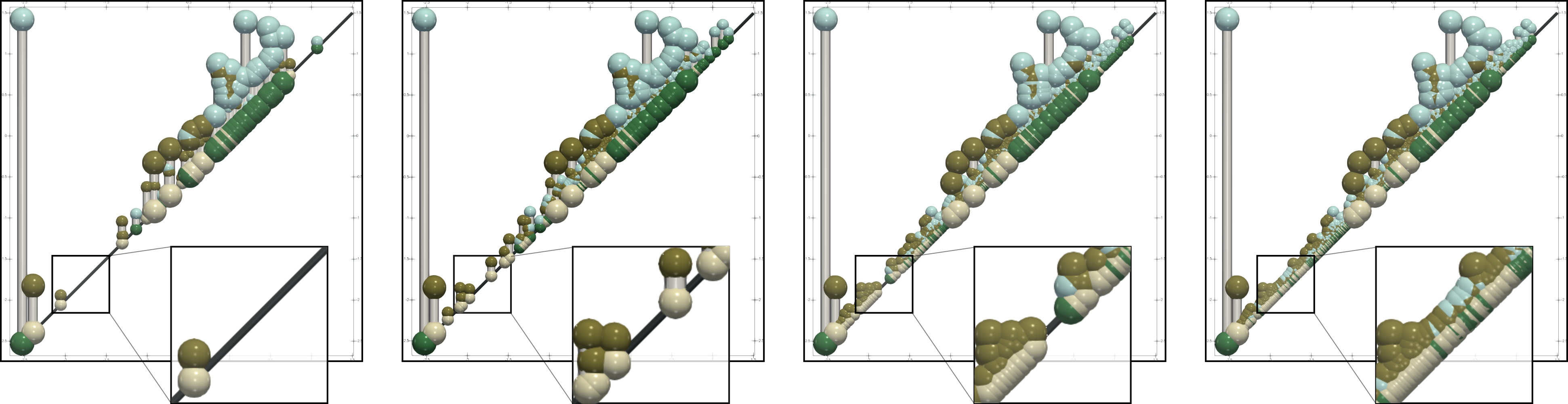}
 \imageCaption{Interrupted Wasserstein barycenters for one cluster of the 
\textit{Sea Surface Height} ensemble with different computation time 
constraints. From left to right : 0.1~s., 1~s., 10~s., and full convergence
\needsVerification{(21 s.)}. 
\vspace{.75ex}
}
  \label{fig_interrupted}
\end{figure}

\removable{As discussed in \autoref{sec_relatedWork}, the \emph{Sinkhorn} 
method \cite{lacombe2018} considers an intermediate heat map representation of 
persistence diagrams and does not explicitly produce a diagram on its output, 
which limits its applicability. Moreover, the geometrical lifting described in 
\autoref{sec_distance}, which is particularly useful in our
applications, is difficult to express in this Eulerian setting, 
where 5-dimensional histograms would be needed, which is impractical. In 
contrast, the \emph{Munkres} and \emph{Auction} approaches produce explicit 
barycenters and optimize the same functional as our approach 
(\autoref{eq_frechet}), which allows direct comparison.}

As predicted, the cubic time complexity of the \emph{Munkres} 
algorithm
makes it impractical for barycenter estimation, as the 
computation completed within 24 hours for only two 
ensembles\julienRevision{(\emph{Vortex 
Street} and \emph{Starting Vortex}), where the diagrams are particularly 
small.}{.}
The \emph{Auction} approach is more practical but still requires up to hours to 
converge for the largest data sets. In contrast, our approach converges in 
sequential in less than 15 minutes at most. The column \emph{Speedup} reports 
the gain obtained with our method against the fastest of the two 
explicit alternatives, 
\emph{Munkres} or \emph{Auction}. For ensembles of 
realistic size, this speedup is about an order of magnitude.
\julienRevision{Our approach can be trivially 
parallelized by running the \emph{Assignment} step (\autoref{alg_final})
in independent threads (\autoref{sec_parallelism}). We implemented this strategy 
with OpenMP and the 
results are reported in \autoref{table_parallel}.}
{As reported in \autoref{table_parallel}, our approach can be trivially 
parallelized with OpenMP by running the \emph{Assignment} step 
(\autoref{alg_final})
in independent threads (\autoref{sec_parallelism}).}
As the size of the input 
diagrams $\diagram(f_i)$ may strongly vary within an ensemble, this trivial 
parallelization may result in load imbalance among the threads, impairing 
parallel efficiency. In practice, this strategy still
provides reasonable speedups, bringing the computation down to a couple of 
minutes at most.

%
%
%
%
%
%
%
%
%
%

\subsection{Barycenter quality}
\autoref{table_energy} compares the Fr\'echet energy (\autoref{eq_frechet}) 
of the converged barycenters for our method and the \emph{Auction barycenter} 
alternative \mixRef \hspace{.25ex}\julienRevision{}{(the Wasserstein distances 
between the results of the two approaches are provided in additional material 
for further details)}.
\julienRevision{In particular, t}{T}he Fr\'echet energy has 
been precisely evaluated with an estimation of Wasserstein distances 
based on the Auction algorithm run until convergence 
(\needsVerification{$\auctionTolerance = 0.01$}).
While 
the actual  values for this energy are not specifically relevant (because of 
various data ranges), the
ratio between the two methods indicates that the local minima approximated by 
both 
approaches are of comparable numerical quality, with a variation of $2\%$ in 
energy at most. 
\autoref{fig_visualComparison} provides a visual comparison of 
the converged Wasserstein barycenters obtained with the \emph{Auction 
barycenter} 
alternative
\mixRef
~ and our method, for one cluster of each of our data sets 
(\autoref{sec_clusteringResults}). This 
figure shows that differences are barely noticeable and only involve 
 pairs with low persistence,
which are often of small interest in the applications.

\autoref{fig_convergence} \julienRevision{details}{compares} the convergence 
rates of the \emph{Auction 
barycenter} 
\mixRef
(red) \julienRevision{and}{to} three variants of our 
framework: without (blue) and with (orange) persistence progressivity and with 
time computation constraints (green\julienRevision{}{, complete 
computations for increasing time constraints}). \julienRevision{For the 
interruptible version 
of our algorithm (green), the energy is reported after each computation,
for increasing time constraints.}{}
\julienRevision{This plot}{It} indicates that our approach based on 
\emph{Price Memorization} and single auction round 
\julienRevision{(Secs. \ref{sec_assignmentReuse}, \ref{sec_partialbidding}, 
blue}
{(blue}) already 
substantially accelerates convergence (the first iteration\julienRevision{after 
initialization}{}, dashed, is performed with a large $\epsilon$ and thus 
induces a high energy). Interestingly, persistence-driven progressivity (orange)
provides the most important gains in convergence speed. 
The number of \emph{Relaxation} iterations 
is larger for our 
approach ($43$, orange) than for the \emph{Auction barycenter} method ($23$, 
red), which 
emphasizes the low computational effort of each of our iterations. 
Finally, when the \emph{Auction barycenter} method completed its first 
\emph{Relaxation} iteration (leftmost red point), our persistence-driven 
progressive algorithm already achieved \needsVerification{80\%} of its
    iterations, resulting in a Fr\'echet energy almost twice smaller.
    The quality of the barycenters obtained with the interruptible 
version of our approach (\autoref{sec_interruptible}) 
is illustrated 
in Figs. \ref{fig:teaser} and \ref{fig_interrupted} for varying time 
constraints. 
As predicted, features of decreasing persistence progressively 
appear in the diagrams, while the most salient features 
are 
accurately represented
for very small constraints, allowing for reliable 
estimations within interactive times (below a second). 


\begin{figure}
\includegraphics[width=\columnwidth]{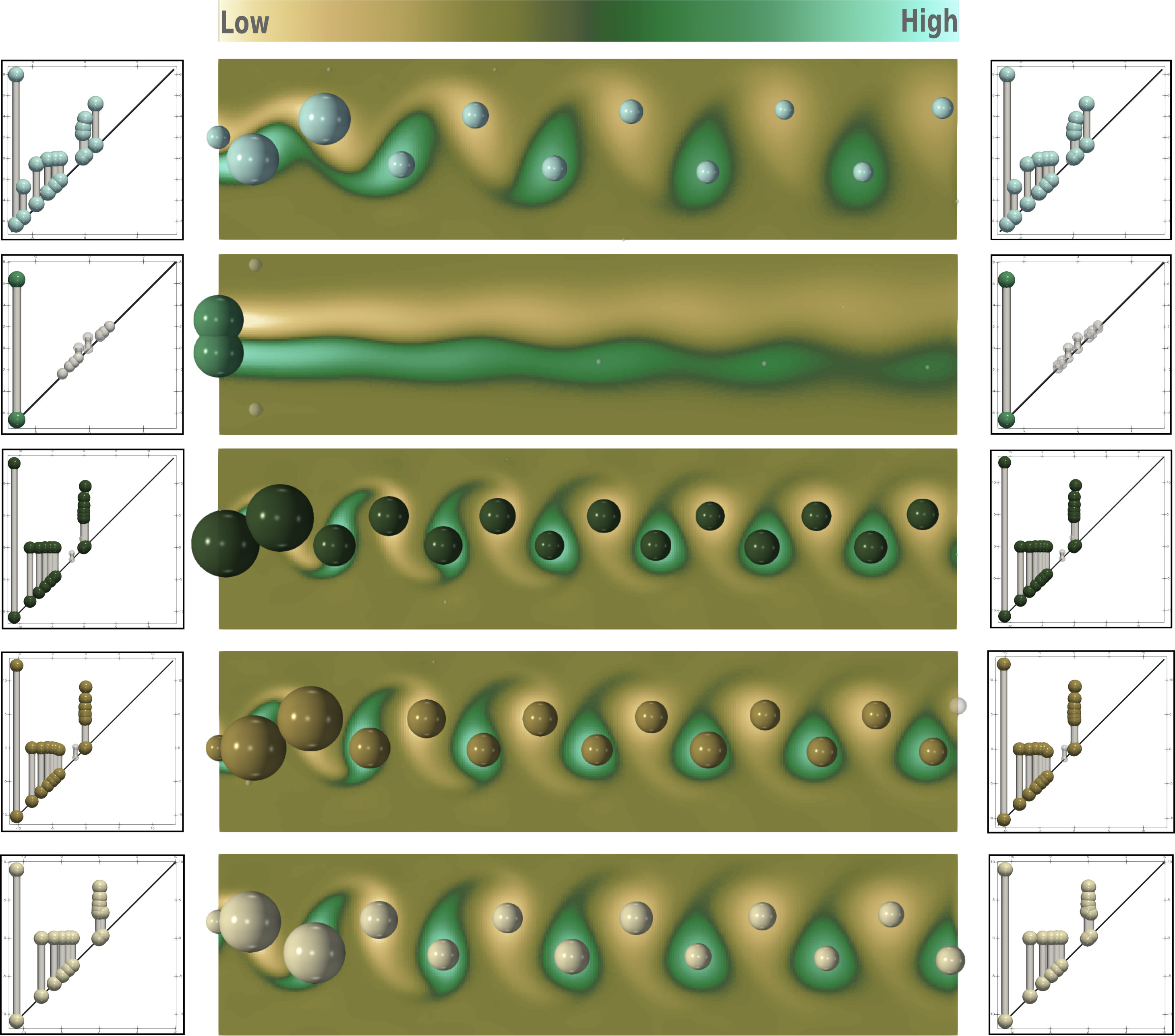}
 \imageCaption{Clusters automatically identified by our topological clustering 
($\timeConstraint$: 10 seconds). From top to bottom: pointwise mean 
of each cluster. 
Left: Centroids computed by our interruptible clustering algorithm. Right: 
Wasserstein barycenters of the clusters, computed by our progressive algorithm 
run until convergence. 
Differences are visually indistinguishable.
Barycenter extrema are scaled in the domain by persistence (spheres).
\todo{embedding seems slightly 
off, use 1-0 and 0-1 for pair embedding.}}
%
%
\label{fig_vortexStreet}
\end{figure}

\subsection{Ensemble visual analysis with Topological Clustering}
\label{sec_clusteringResults}
\begin{figure}
\includegraphics[width=\columnwidth]{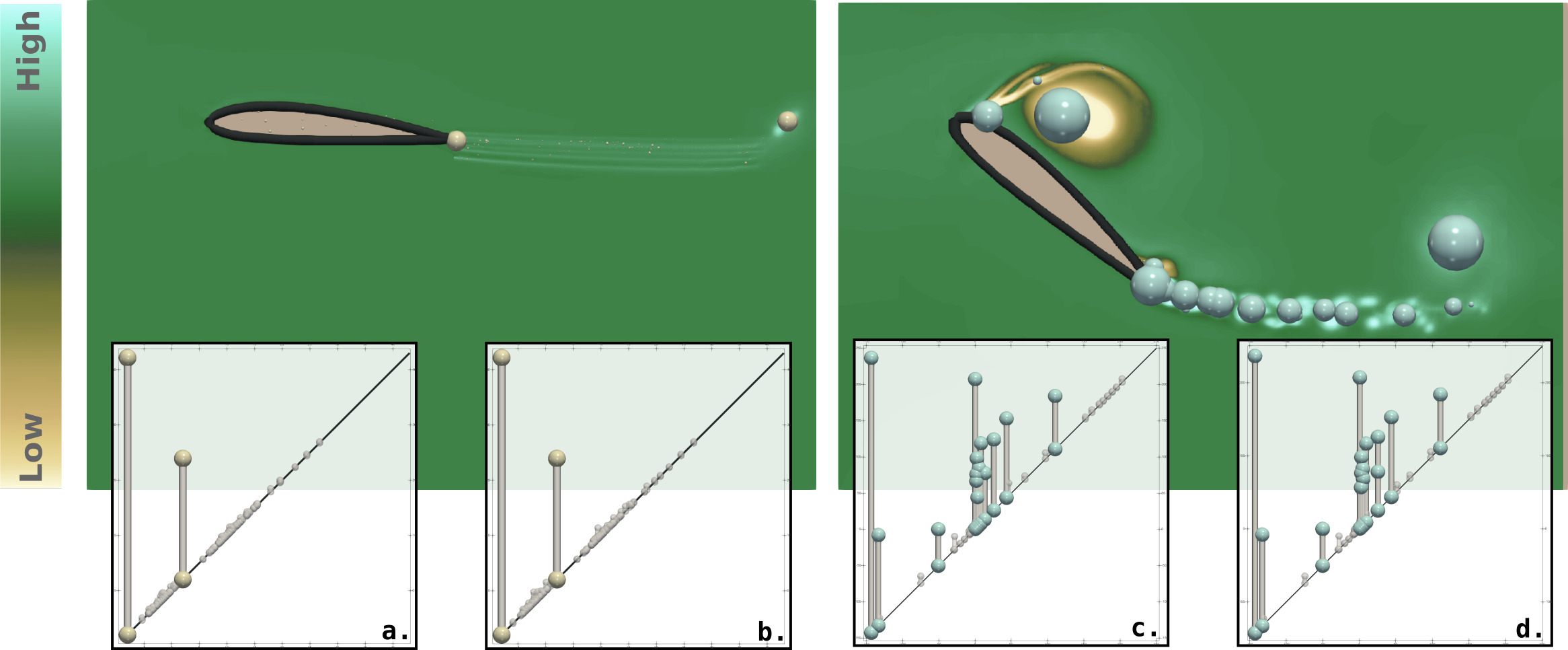}
 \imageCaption{Clusters automatically identified by our topological clustering 
($\timeConstraint: 10 s.$).
Left insets \julienRevision{}{(a, c)}: Centroids computed by our interruptible 
clustering 
algorithm.
Right insets \julienRevision{}{(b, d)}: 
Wasserstein barycenters of the clusters, computed by our progressive algorithm 
run until convergence. 
Differences are visually indistinguishable. 
\julienRevision{
Top: pointwise mean of each cluster.
Barycenter extrema are scaled in the domain by 
persistence (spheres).}
{Top: pointwise mean of each cluster, with barycenter extrema scaled by 
persistence (spheres).}
\todo{Same as above, embedding slightly off.}}
\label{fig_startingVortex}
\end{figure}

In the following, we 
systematically set a time constraint $\timeConstraint$  of 10 seconds. 
To facilitate the reading of the diagrams, each 
pair with a  persistence smaller than $10\%$ of the function range 
is shown in transparent white, to help visually discriminate salient features 
from noise.
\autoref{fig_gaussians} shows the 
clustering of the \emph{Gaussians} ensemble by our approach. This synthetic 
ensemble
exemplifies the motivation for the geometrical lifting 
\julienRevision{introduced in \autoref{sec_distance}}{(\autoref{sec_distance})}.
The 
first and third clusters 
both contain a single Gaussian, 
resulting in diagrams with a single persistent feature, but located in 
drastically different areas of the domain $\domain$. 
Thus, the diagrams of these two clusters would be indistinguishable for the 
clustering algorithm if geometrical lifting was not considered. If feature 
location is important for the application, our approach can be adjusted thanks 
to geometrical lifting (\autoref{sec_distance}). For the \emph{Gaussians} 
ensemble, this makes our clustering approach compute the correct clustering. 
Moreover, taking the geometry of the critical points into account allows us to 
represent in $\domain$ the extrema involved in the Wasserstein barycenters 
(spheres, scaled by persistence, \autoref{fig_gaussians}) which allows user to 
have a visual feedback in the domain of the features representative of the set 
of scalar fields. Geometrical lifting is particularly important in applications 
where feature location bears a meaning, such as the Isabel ensemble 
(\autoref{fig:teaser}(f)). For this example, our clustering algorithm with 
geometrical lifting automatically identifies the right clusters, corresponding 
to the three states of the hurricane (formation, drift and landfall). For the 
remaining examples, geometrical lifting was not necessary 
($\geometricLifting = 0$).
For the \emph{Vortex Street} ensemble (\autoref{fig_vortexStreet}), our 
approach manages to automatically identify the correct clusters, 
precisely corresponding to the distinct viscosity regimes found in the 
ensemble. Note that the centroids computed by our topological clustering 
algorithm with a time constraint of 10 seconds (left) are visually 
indistinguishable from the Wasserstein barycenters of each cluster,
 computed after the fact with our progressive algorithm run until 
convergence (right). This indicates that the centroids provided by our 
topological clustering are reliable and can be used to visually represent the 
features of interest in the ensemble. 
In particular, for the \emph{Vortex 
Street} example, these centroids enable the clear identification of the number 
and salience of the vortices: 
pairs which align horizontally and vertically respectively denote  minima 
and maxima of flow vorticity, which respectively correspond to clockwise and 
counterclockwise vortices. 
\autoref{fig_startingVortex} presents our results on the \emph{Starting Vortex}, 
where our approach also automatically identifies the correct 
clustering, corresponding to two wing configurations. In this example, the 
difference in turbulence (number and strength of vortices) can be directly 
and visually read from the centroids returned by our algorithm (insets).
Finally, 
\autoref{fig_seaSurfaceHeight} shows our results for the \emph{Sea Surface 
Height}, where our topological clustering automatically identifies four 
clusters, corresponding to the four seasons: 
\needsVerification{winter (top left), spring (top 
right), summer (bottom left), fall (bottom right).} 
As shown in the insets, each season leads to a visually distinct centroid 
diagram.
In this example, 
as diagrams are larger, differences between the interrupted centroids (left) 
and the converged  barycenters (right) become noticeable. However, 
these differences only involve 
pairs of small persistence, 
whose contribution to the final clustering reveal negligible in practice. 

\julienRevision{\EDIT{}{
Comparing with the approach from Favelier et al. 
\cite{favelier2018},
on the exact same data, we obtain similar qualitative results, with correct
clusterings found on all data-sets. Performance-wise, our time-constrained
approach finds the correct clustering within ten seconds for a large
 ensemble data-set such as \textit{Sea Surface
Height} (in addition to 35 seconds to compute the persistence diagrams) against several hundreds of seconds for the parallel
persistence atlas approach. 
}}
{Overall, our approach provides the same clustering results than 
Favelier et al. \cite{favelier2018}: the returned clusterings are correct for 
both approaches, for all of the above data sets. However, 
once the input persistence diagrams are available,
our algorithm computes 
within 
a time constraint of ten seconds only,
while the approach by Favelier et al. requires up 
to hundreds of seconds (on the same hardware) to compute intermediate 
representations (\emph{Persistence Maps}) which are not needed in our 
work.}

\subsection{Limitations}
\begin{figure}
\includegraphics[width=\columnwidth]{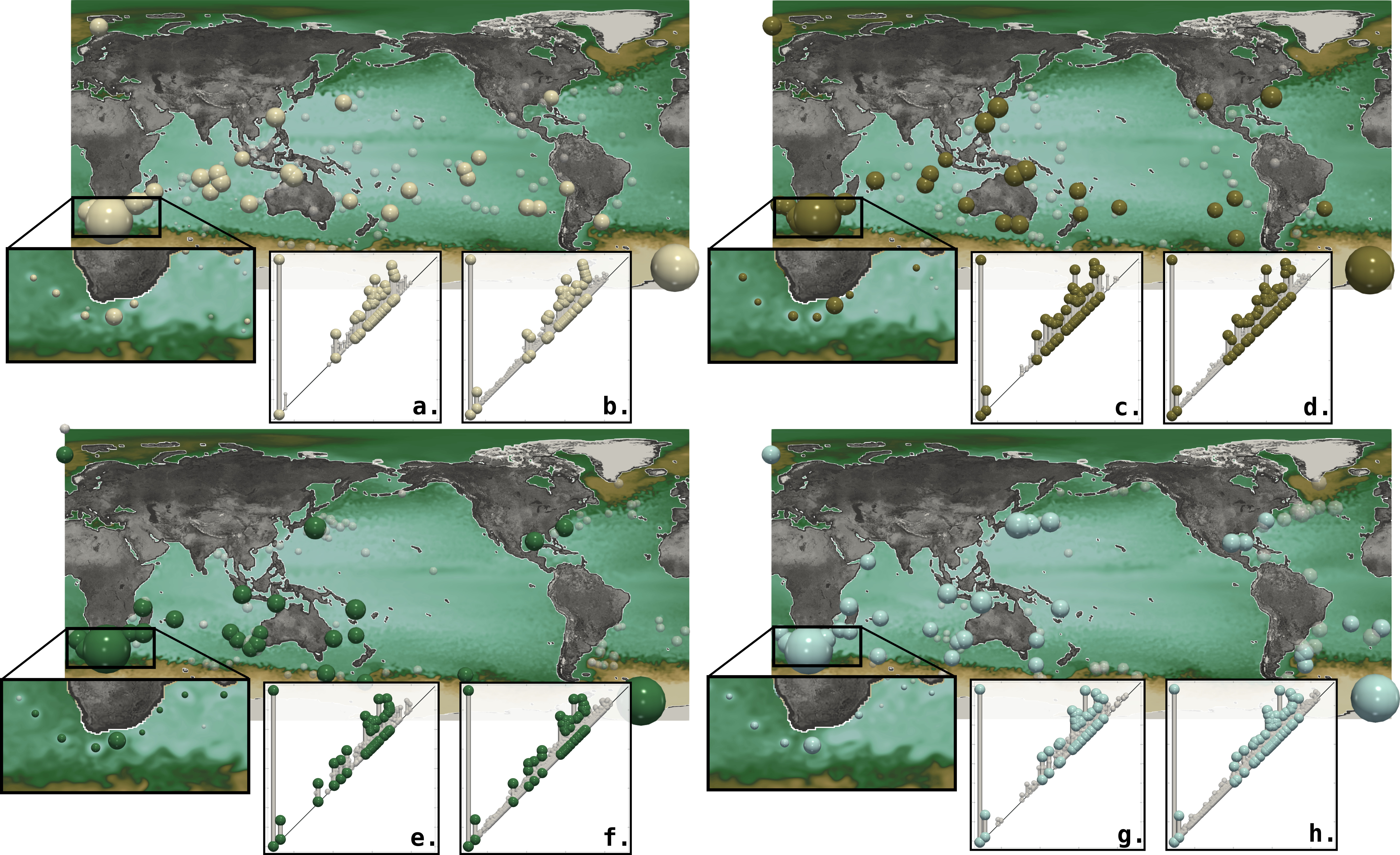}
 \imageCaption{Clusters automatically identified by our 
topological clustering 
 ($\timeConstraint: 10 s.$).
\julienRevision{From left to right, top to bottom:
pointwise mean of each cluster. 
Barycenter extrema
are embedded in the domain with 
spheres scaled by persistence.}
{From left to right, top to bottom: pointwise mean of each cluster with 
barycenter extrema scaled by persistence (spheres).}
Left insets \julienRevision{}{(a, c, e, g)}: Centroids computed by our 
interruptible clustering algorithm. 
Right insets \julienRevision{}{(b, d, f, h)}: 
Wasserstein barycenters of the clusters,
computed by our progressive algorithm 
run until convergence. 
\todo{Same as before, embedding slightly off. there's a vortex on the 
US-mexico wall (top right).}
\todo{What do the clusters correspond to? top-right: winter, bottom left: 
spring, top left: summer: bottom right: fall}}
 \label{fig_seaSurfaceHeight}
\end{figure}

In our experiments, we focused on persistence diagrams which only involve 
extrema, as these often directly translate into features of interest in the 
applications. Although our approach can consider other types of persistence 
pairs (\emph{e.g.} saddle-saddle pairs in 3D), from our 
experience, the interpretation of these structures 
is not obvious in practice and future work is needed to improve the 
understanding of these 
pairs 
in the applications.
Thanks to the assignments computed by our algorithm, the extrema of the output 
barycenter can be embedded in the original domain (\autoref{fig_gaussians} 
to \ref{fig_seaSurfaceHeight}). However, in practice a 
given barycenter extremum can be potentially assigned with 
extrema which are distant from each other in the ensemble members, resulting in 
its placement at an in-between location which may not be relevant for the 
application.
\needsVerification{
Regarding the Fr\'echet energy,
our experiments confirm the proximity of our approximated 
barycenters to  \julienRevision{}{actual} local minima 
(\autoref{fig_visualComparison}\julienRevision{}{, \autoref{table_energy}}). 
\julienRevision{}{However, 
theoretical proximity 
bounds to these minima are difficult to formulate and
we leave this for future work.} \julienRevision{However, a}{Also, a}s it is the 
case 
for the original 
algorithm by Turner et al. \cite{Turner2014}, there is no guarantee that 
\julienRevision{this 
solution is a global minimizer}{our solutions are global minimizers}.} 
For the clustering, we observed that the initialization of the 
$k$-means algorithm had a major impact on its outcome but we found that 
the \emph{$k$-means++} heuristic  \cite{celebi13} provided 
excellent results in practice. Finally, when the geometrical location of 
features in the domain has a meaning for the applications, 
the geometrical lifting coefficient 
(\autoref{sec_distance}) must 
be manually adjusted by the user on a per application basis, which involves a 
trial and error process. However, our interruptible approach greatly helps in 
this process, as users can perform such adjustments at interactive rates.



\section{Conclusion}

In this paper, we presented an algorithm for the progressive approximation of 
Wasserstein barycenters of Persistence diagrams, with applications to the 
visual analysis of ensemble data. 
\julienRevision{\EDIT{}{Our approach takes advantage of previous results on the 
computation of
Fréchet means for persistence diagrams \cite{Turner2014}, and revisits the fast
approximation of Wasserstein distance \cite{Kerber2016} to induce 
progressivity.}}
{Our approach revisits efficient algorithms for Wasserstein 
distance approximation \cite{Bertsekas81, Kerber2016} in order to specifically
extend previous work on barycenter estimation \cite{Turner2014}.}
Our experiments showed that our strategy drastically
accelerates convergence and 
reported an order of magnitude speedup against 
previous work,
while providing barycenters 
which are quantitatively \julienRevision{comparable and visually similar.}
{and visually comparable.}
The progressivity of our 
approach 
\julienRevision{allowed}{allows}
for the definition of an \emph{interruptible} algorithm, 
enabling the 
estimation of reliable barycenters within interactive times. We presented an 
application 
to \julienRevision{the topological clustering of ensemble data,}
{ensemble data clustering,}
where 
 the obtained centroid diagrams provided key visual insights about the 
global feature trends of the ensemble.

A natural direction for future work is the extension of our framework to other 
topological abstractions, such as Reeb graphs or Morse-Smale complexes. 
However, the question of defining a relevant, and importantly, computable 
metric between these objects 
is still an active research 
debate.
\needsVerification{Our framework provides only approximations of Wasserstein 
barycenters. In the future, it would be useful to study the convergence of these 
approximations from a theoretical point of view.}
Although we have focused on scientific visualization applications, our 
framework can be used mostly as-is for persistence diagrams of more general 
data, such as filtrations of high dimensional point clouds. 
\julienRevision{}{In that context, other applications than clustering will also 
be investigated.}
Moreover, we 
believe our 
progressive strategy for Wasserstein barycenters can also be used for more 
general inputs than persistence diagrams, such as generic point clouds, as long 
as an importance measure substituting persistence is available, which would 
significantly enlarge the spectrum of applications of the ideas presented in 
this paper.



%

\acknowledgments{
\scriptsize{
The authors would like to thank T. Lacombe, M. Cuturi and S. Oudot for sharing 
an implementation of their approach \cite{lacombe2018}. We also thank the 
reviewers for their thoughtful remarks and suggestions.
This work is partially supported by the European Commission grant 
H2020-FETHPC-2017 ``VESTEC'' (ref. 800904). 
}}

\clearpage

\bibliographystyle{abbrv}

\bibliography{pdbarycenter}

\end{document}